\documentclass[cleanhead, cleanfoot,10pt,twocolumn]{asme2e}

\usepackage{empheq, amssymb}
\usepackage{lmodern}
\usepackage{amsmath}
\usepackage{amsfonts}
\usepackage{amssymb}
\usepackage{bm} % bold for greek symbols
\usepackage{txfonts}
\usepackage[T1]{fontenc}
\usepackage[utf8]{inputenc} % accept different input encodings
\usepackage[dvipsnames]{xcolor} % defines colors
\usepackage[american]{babel} %  culturally-determined typographical rules
\usepackage{graphicx}
\usepackage{epstopdf}
\usepackage[noadjust]{cite} % grouped citations
\usepackage{ltxcmds}
\usepackage{mathtools}
\usepackage{subcaption}% for subfigures
\usepackage{multirow} %for tables 
\usepackage{blindtext}
\usepackage{hyperref}
\usepackage{enumitem}
\usepackage{ifthen}
\usepackage{etoolbox}
\usepackage{cancel}
\usepackage{empheq}
\usepackage{fontawesome}
\usepackage{soul}
\usepackage{textcomp}
\usepackage{lipsum}

\definecolor{light-gray}{gray}{0.4}
\definecolor{box-gray}{gray}{1}

\hypersetup{
    unicode=false,          % non-Latin characters in Acrobat’s bookmarks
    pdftoolbar=true,        % show Acrobat’s toolbar?
    pdfmenubar=true,        % show Acrobat’s menu?
    pdffitwindow=false,     % window fit to page when opened
    pdfstartview={FitV},    % fits the width of the page to the window
    pdftitle={An Overview of Uncertain Control Co-design Formulations},    % title
    pdfauthor={Saeed Azad and Daniel R. Herber},     % author
    pdfsubject={},   % subject of the document
    pdfkeywords = {}, % list of keywords
    pdfnewwindow=true,      % links in new window
    colorlinks=true,
    allcolors=blue
}
\usepackage{nomencl}
\renewcommand\nomgroup[1]{%
  \item[\bfseries
  \ifstrequal{#1}{V}{ Key Variables}{%
  \ifstrequal{#1}{B}{ Subscripts}{%
  \ifstrequal{#1}{P}{ Notation}{%
  \ifstrequal{#1}{A}{ Acronyms}{}}}}]
}

\makenomenclature



\usepackage{dblfloatfix}
\usepackage{float}

\definecolor{block-gray}{gray}{0.95}

\usepackage[framemethod=TikZ]{mdframed}

\usepackage{xpatch}

\makeatletter
\xpatchcmd{\endmdframed}
  {\aftergroup\endmdf@trivlist\color@endgroup}
  {\endmdf@trivlist\color@endgroup\@doendpe}
  {}{}
\makeatother

\usepackage{fontawesome}

\usepackage{titlesec}

\newcommand{\doi}[1]{{doi:~\href{https://doi.org/#1}{\nolinkurl{#1}}}\rmFullStop}

\newcommand*{\rmFullStop}{\rmifnextchar{.}{}{}}

\makeatletter
\newcommand{\rmifnextchar}[3]{%
  \begingroup
  \ltx@LocToksA{\endgroup#2}%
  \ltx@LocToksB{\endgroup#3}%
  \ltx@ifnextchar{#1}{%
    \def\next{\the\ltx@LocToksA}%
    \afterassignment\next
    \let\scratch= %
  }{%
    \the\ltx@LocToksB
  }%
}
\makeatother

\usepackage{colortbl}

\definecolor{light-gray}{gray}{0.6}

\newcommand{\xsection}[1]{\section[#1]{\MakeUppercase{#1}}}
\usepackage{accents}

\usepackage{xcolor}% http://ctan.org/pkg/xcolor
\newcommand{\parm}{\mathord{\color{black!33}\bullet}}%

\usepackage{nomencl}
\makenomenclature

\usepackage{mathrsfs}
\usepackage{euscript}
\usepackage{mathtools}

\usepackage{caption}
\usepackage{subcaption}
\usepackage{tikz}
\usepackage[absolute,overlay]{textpos}
\allowdisplaybreaks
\usepackage{tikz}
\usetikzlibrary{trees, backgrounds, calc, positioning, fit}
\title{An Overview of Uncertain Control Co-Design Formulations}
\author{Saeed~Azad\thanks{Corresponding author, \texttt{\href{mailto:saeed.azad@colostate.edu}{saeed.azad@colostate.edu}}} 
\affiliation{
Postdoctoral Fellow\\
Department of Systems Engineering \\
Colorado State University\\
Fort Collins, CO 80523 \\
Email:~\texttt{\href{mailto:saeed.azad@colostate.edu}{saeed.azad@colostate.edu}}\\ 
}
}
\author{Daniel~R.~Herber
\affiliation{
Assistant Professor\\
Department of Systems Engineering \\
Colorado State University \\
Fort Collins, CO 80523 \\
Email:~\texttt{\href{mailto:daniel.herber@colostate.edu}{daniel.herber@colostate.edu}}
}
}

\begin{document}

\setlength{\parskip}{0pt}
\setlength{\parsep}{0pt}
\setlength{\headsep}{0pt}
\setlength{\topsep}{0pt}

% equations
\abovedisplayshortskip=3pt
\belowdisplayshortskip=3pt
\abovedisplayskip=3pt
\belowdisplayskip=3pt

\titlespacing*{\section}{0pt}{18pt plus 1pt minus 1pt}{3pt plus 0.5pt minus 0.5pt}
\titlespacing*{\subsection}{0pt}{9pt plus 1pt minus 0.5pt}{1pt plus 0.5pt minus 0.5pt}
\titlespacing*{\subsubsection}{0pt}{9pt plus 1pt minus 0.5pt}{1pt plus 0.5pt minus 0.5pt}

\maketitle

%-----------------abstract----------------%
\begin{abstract}\noindent
\textit{This article explores various uncertain control co-design (UCCD) problem formulations. While previous work offers formulations that are method-dependent and limited to only a handful of uncertainties (often from one discipline), effective application of UCCD to real-world dynamic systems requires a thorough understanding of uncertainties and how their impact can be captured.
Since the first step is defining the UCCD problem of interest, this article aims at addressing some of the limitations of the current literature by identifying possible sources of uncertainties in a general UCCD context and then formalizing ways in which their impact is captured through problem formulation alone (without having to immediately resort to specific solution strategies).
We first develop and then discuss a generalized UCCD formulation that can capture uncertainty representations presented in this article. 
Issues such as the treatment of the objective function, the challenge of the analysis-type equality constraints, and various formulations for inequality constraints are discussed.
Then, more specialized problem formulations such as stochastic in expectation, stochastic chance-constrained, probabilistic robust, worst-case robust, fuzzy expected value, and possibilistic chance-constrained UCCD formulations are presented.
Key concepts from these formulations, along with insights from closely-related fields, such as robust and stochastic control theory, are discussed, and future research directions are identified.}
\end{abstract}

\vspace{1ex}

%--------------Introduction-----------------%
\xsection{Introduction}\label{sec:introduction}

% new paragraph
With the ever-growing complexity and integrated nature of dynamic engineering systems, the need for effective control co-design (CCD) strategies, i.e., integrated consideration of the physical and control system design, is ever present~\cite{garcia2019control, allison2014special}.
When investigating a CCD problem, it is often the case that some of its elements (e.g., inputs, model parameters, and/or some aspects of system dynamics) are inherently uncertain or not entirely known.
In this paper, we refer to both of these characteristics as uncertainty.
Overlooking the impact of uncertainties in CCD may result in solutions that are no longer effective in realistic scenarios. 

% new paragraph
These uncertainties may stem from multiple sources and affect various elements of the CCD activity. For example:
\begin{enumerate}[topsep=0pt,itemsep=-1ex,partopsep=1ex,parsep=1ex,label=$\bullet$]
\item The noise acting through the control channel transforms the deterministic control trajectories into stochastic ones
\item Plant optimization variables may be uncertain due to imperfect manufacturing processes, measurement errors, and mass production of components
\item Uncertain problem data (such as wind speeds, wave energy densities, earthquake loads, and material properties) may also affect various elements of the problem
\item Fidelity of the dynamic model (i.e.,~unmodeled or neglected dynamics) may be another source of uncertainty that often arises as a trade-off between model simplicity and accuracy
\end{enumerate}
\noindent 
All of these uncertainties may propagate through the dynamic system and transform the states into uncertain trajectories.
Consequently, such uncertainties transform the CCD problem into an \textit{uncertain control co-design} (UCCD) problem.
Even before attempting to solve such problems, a necessary step is to identify ways in which the impact of such uncertainties can be mathematically captured in an optimization formulation context.    
Therefore, it is critical to establish and understand various possible UCCD problem formulations.

% new paragraph
This paper aims to identify the sources of uncertainties and formalize their inclusion in various UCCD formulations.
This contribution is motivated by the fact that, currently, uncertainty quantification is reasonably well understood in specific control and plant design optimization communities~\cite{du2002efficient,ba2006new,yang2011robust,dullerud2013course,aastrom2012introduction}.
However, current UCCD studies in the literature generally suffer from the lack of a holistic view towards uncertainties, focusing on specific uncertainties, often motivated by a particular solution technique~\cite{azad2020single,azad2020robust,nash2021robust,cui2020comparative, behtash2021reliability}. 
Therefore, the distinction between various UCCD problem formulations is rarely discussed. 

% new paragraph
In this article, we present an initial effort at a generalized UCCD problem formulation. 
Various problem elements, including the optimization variables, objective function, equality and inequality constraints, and relevant concepts such as risk, are discussed.
Next, we transition towards specialized formulations that are motivated by concepts from stochastic programming~\cite{ruszczynski2003stochastic,powell2019unified}, robust optimization~\cite{beyer2007robust,gorissen2015practical,bertsimas2011theory}, and fuzzy programming~\cite{zadeh1996fuzzy,zadeh1978fuzzy,liu2009theory}. 
These formulations provide the necessary framework for the development and widespread adoption of UCCD formulations in order to meet the ever-increasing demands on performance, robustness, and reliability of real-world dynamic systems. 
For more information on the implementations and specific engineering applications, readers are encouraged to consult the references provided within the article.
Detailed reviews on uncertainty-based approaches for various engineering applications, such as aerospace vehicles, distributed energy systems, motion planning, process scheduling, power systems, building energy assessment, and wind power forecasting, can be found in Refs.~\cite{yao2011review, mavromatidis2018review, dadkhah2012survey, li2008process, aien2016comprehensive, tian2018review, yan2015reviews}, respectively.

% new paragraph
The remainder of this article is organized as follows: 
Sec.~\ref{sec:2} describes the deterministic CCD problem formulation and various representations of uncertainty; 
Sec.~\ref{sec:3} provides a mathematical foundation for a general UCCD problem formulation; 
Sec.~\ref{sec:4} describes some of the specialized UCCD formulations that are inspired by stochastic, worst-cast robust, and fuzzy programming frameworks, including stochastic in expectation UCCD, stochastic chance-constrained UCCD, probabilistic robust UCCD, worst-case robust UCCD, fuzzy expected value UCCD, and possibilistic chance-constrained UCCD;
and Sec.~\ref{sec:5} discusses several more specific topics in the context of UCCD.
Finally, Sec.~\ref{sec:conclusion} presents the conclusions.

%--------------Section 2-----------------%
\section{Uncertainty Representations in UCCD} \label{sec:2}
In this section, the deterministic CCD, which is a special case of UCCD formulation, is introduced.
For mathematical clarity, we define sets associated with both time-dependent and time-independent deterministic and uncertain variables.
This section also introduces three distinct ways to represent uncertainties in UCCD context: stochastic, crisp, and possibilistic.    
\subsection{Deterministic CCD}
\label{subsec:DCCD}
We begin by introducing the nominal continuous-time, deterministic, all-at-once (AAO), simultaneous, CCD problem:
\begin{subequations}
 \label{Eqn:DCCD}
 \begin{align}
 \underset{\bm{u}, \bm{\xi}, \bm{p}}{\textrm{minimize:}} \quad & o=\displaystyle\int_{t_0}^{t_f}  \ell(t,\bm{u},\bm{\xi}, \bm{p}, \bm{d})\,\mathrm{d}t + m (\bm{p}, \bm{\xi}_{0},\bm{\xi}_{f}, \bm{d})\label{Eqn:DCCD_obj}\\
 \textrm{subject to:} 
 \quad & \bm{g}(t,\bm{u},\bm{\xi}, \bm{p}, \bm{\xi}_{0}, \bm{\xi}_{f},\bm{d})\leq \bm{0}\label{Eqn:DCCD_ineq}\\
 \quad & \bm{h}(t,\bm{u},\bm{\xi}, \bm{p}, \bm{\xi}_{0}, \bm{\xi}_{f}, \bm{d})=\bm{0} \label{Eqn:DCCD_eq}  \\
 \quad & \dot{\bm{\xi}} - \bm{f}(t,\bm{u},\bm{\xi}, \bm{p},\bm{\xi}_{0}, \bm{\xi}_{f}, \bm{d}) = \bm{0} \label{Eqn:DCCD_dyn}\\ 
 \textrm{where:}
 \quad & \bm{\xi}(t_0) = \bm{\xi}_{0},~\bm{\xi}(t_f) = \bm{\xi}_{f},~\bm{u}(t) = \bm{u},~\bm{\xi}(t) = \bm{\xi} \label{Eqn:DCCD_ini} \\
 & \bm{d}(t) = \bm{d} \nonumber
 \end{align}
 \end{subequations}
\noindent
where $t \in [t_0, t_f]$ is the time horizon, $\{ \bm{u}, \bm{\xi}, \bm{p} \}$ are the collection of optimization variables including the open-loop control trajectories $\bm{u}(t) \in \mathbb{R}^{n_u}$, state trajectories $\bm{\xi}(t) \in \mathbb{R}^{n_s}$, and the vector of time-independent optimization variables $\bm{p} \in \mathbb{R}^{n_p}$, respectively.
Note that $\bm{p}$ may entail plant optimization variables $\bm{p}_{p}$, and/or time-independent control optimization variables \cite{fathy2003nested, herber2017unified} (i.e.,~gains $\bm{p}_{c}$, such that $\bm{p} = [\bm{p}_{p}, \bm{p}_c]$).
The objective function $o(\cdot)$ is composed of the Lagrange term $\ell(\cdot)$ and the Mayer term $m(\cdot)$.
The vectors of inequality and equality constraints are described by $\bm{g}(\cdot)$ and $\bm{h}(\cdot)$, respectively.
The transition or state derivative function $\bm{f}(\cdot)$ describes the evolution of the system through time in terms of a set of ordinary differential equations (ODEs). 
All of the data associated with the problem formulation is represented through $\bm{d} \in \mathbb{R}^{n_{\bm{d}}}$.
This data, which may be time-dependent or time-independent, includes information such as problem constants, environmental signals, initial/final times, etc.

% new paragraph
In the remainder of this article, we assume that constraints associated with the initial and final conditions $\{\bm{\xi}_{0}, \bm{\xi}_{f}\}$ are already included in $\bm{h}(\cdot)$ or $\bm{g}(\cdot)$. 
In addition, we will often drop the explicit dependence on $t$ from time-dependent quantities such as control and state trajectories, as well as the problem data.
For more details on deterministic CCD, the readers are referred to Refs.~\cite{allison2014special, herber2019nested}. 

\subsection{Representation of Uncertainties}
\label{subsec:RoU}
The first step in accounting for uncertainties in a UCCD problem is the representation of input and model uncertainties.
In the risk assessment context, these uncertainties are either aleatory (irreducible) or epistemic (reducible) \cite{ghanem2003stochastic}.
Aleatory uncertainty is associated with the inherent irregularity of the phenomenon, while epistemic uncertainty is associated with the lack of knowledge.
Accordingly, acquiring more knowledge cannot reduce aleatory uncertainties, but it can reduce epistemic uncertainties.
In fact, epistemic uncertainty captures the analyst's confidence in the model by quantifying their degree of belief in how well the model represents the reality \cite{zio2013literature}. 
As an example, consider the uncertainty in plant optimization variables due to imperfect manufacturing processes.  
Noting that manufacturing processes remain imperfect even when improved, this uncertainty is intrinsically aleatory or irreducible.
This is because acquiring more knowledge cannot reduce this uncertainty (no two plants are identical).
However, the uncertainty in the true probability distribution of plant optimization variables can be reduced by acquiring more knowledge (observations). 
Therefore, this is an epistemic-type uncertainty.
Another example of aleatory uncertainty is randomness in material properties or flipping a biased coin. 
However, our belief in the probabilistic and distributional information of such a phenomenon is epistemic.

% new paragraph
Conventionally, these two types of uncertainty are segregated in a nested algorithm, with aleatory analysis in the inner loop and epistemic analysis on the outer loop~\cite{hofer2002approximate}.
While this allows for the simple separation and tracking of each type of uncertainty, a uniform treatment of aleatory and epistemic uncertainties has been implemented in the literature \cite{wang2021extended} and assumed in this article.
It is important to note that information scarcity on epistemic uncertainties may render the output probabilistic information impractical.
Therefore, when complete distributional information is available, it should be integrated into the UCCD problem.
However, in the case of incomplete and limited information, methods associated with epistemic uncertainties, such as fuzzy programming, are generally preferred. 

% new paragraph
Elements in a UCCD problem formulation may be deterministic or uncertain.
In this article, the notation $\tilde{\parm}$ is used to distinguish uncertain quantities from deterministic ones. Stochastic processes are distinguished with a time argument $\tilde{\parm}(t)$.
We note that while a common assumption is the discretization of the time dimension, nearly all aspects of the continuous definitions of the uncertainties could be considered for discrete realizations at particular time instances only (as is often done for realizability through information availability updates and solution strategies implementations).
For the sake of consistency, all of the formulations in the article are presented in continuous time.
In addition, properties such time-dependence between the various time-dependent signals, such as cross-correlation, are not assumed.   
Such properties might be used to define uncertain quantities and their relations to aspects of a UCCD problem.

\begin{figure}[t]
\centering
\begin{subfigure}[t]{0.5\columnwidth}
\centering
\includegraphics[scale=0.32]{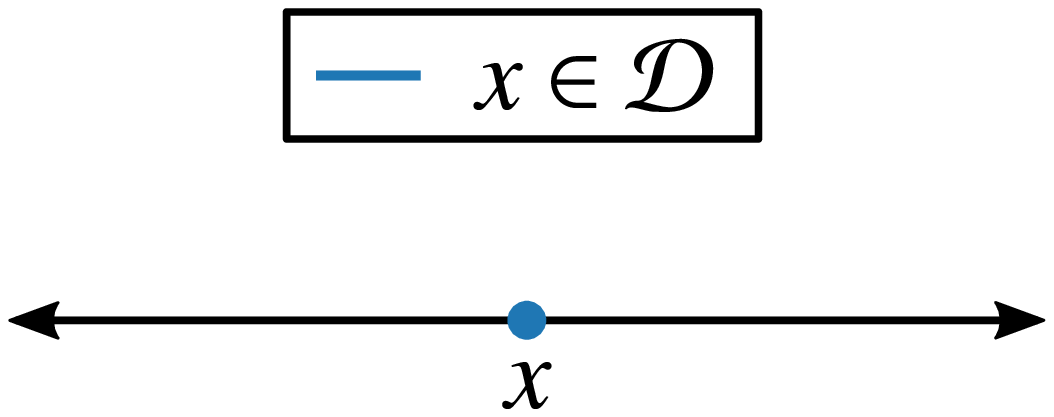}
\caption{Time-independent deterministic.}
\label{subfig:Vd}
\end{subfigure}%
\hspace{0.01\columnwidth}%
\begin{subfigure}[t]{0.49\columnwidth}
\centering
\includegraphics[scale=0.32]{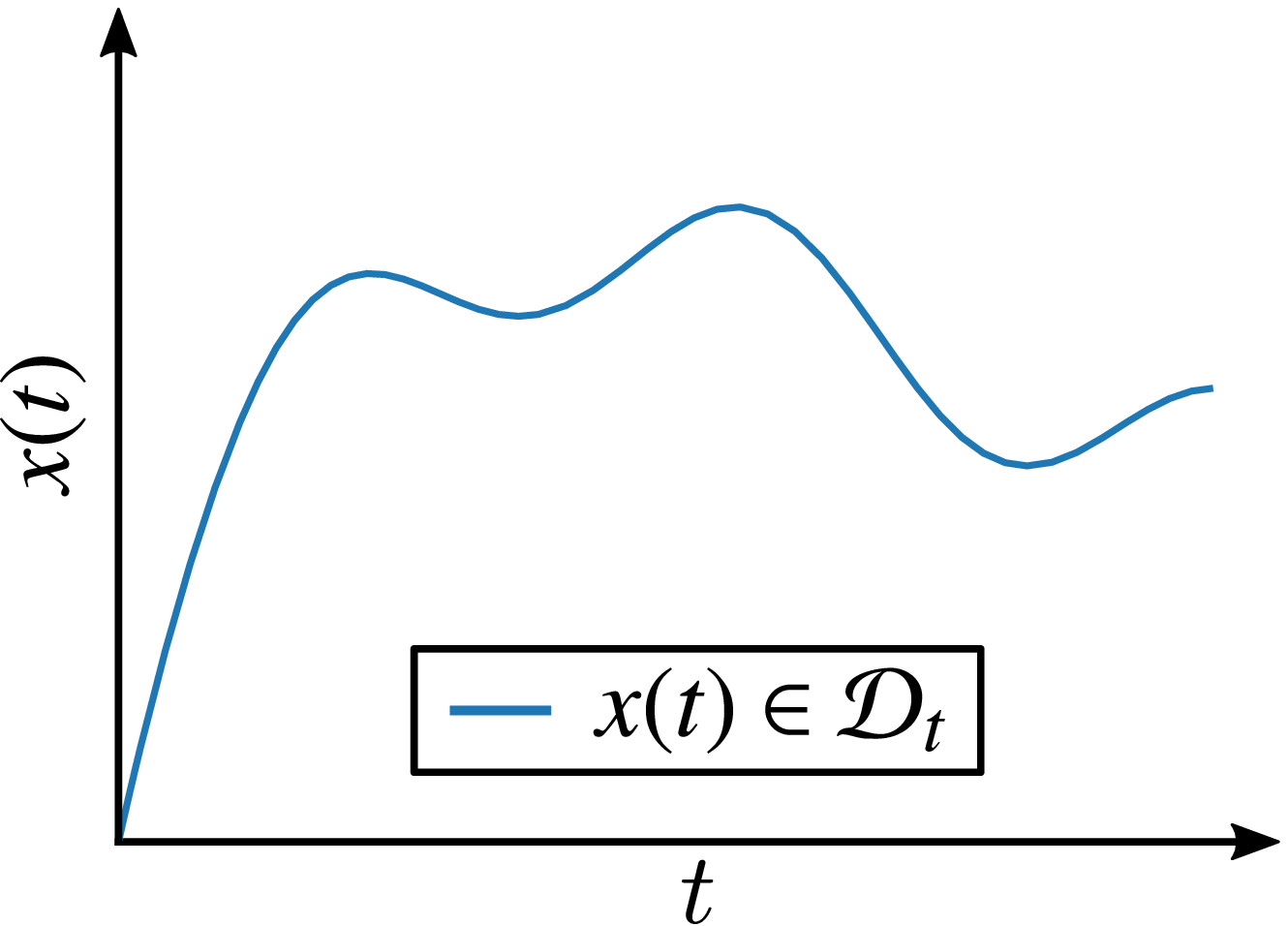}
\caption{Time-dependent deterministic.}
\label{subfig:Td}
\end{subfigure}%
\hspace{0.015\textwidth}%

\begin{subfigure}[t]{0.495\columnwidth}
\centering
\includegraphics[scale=0.32]{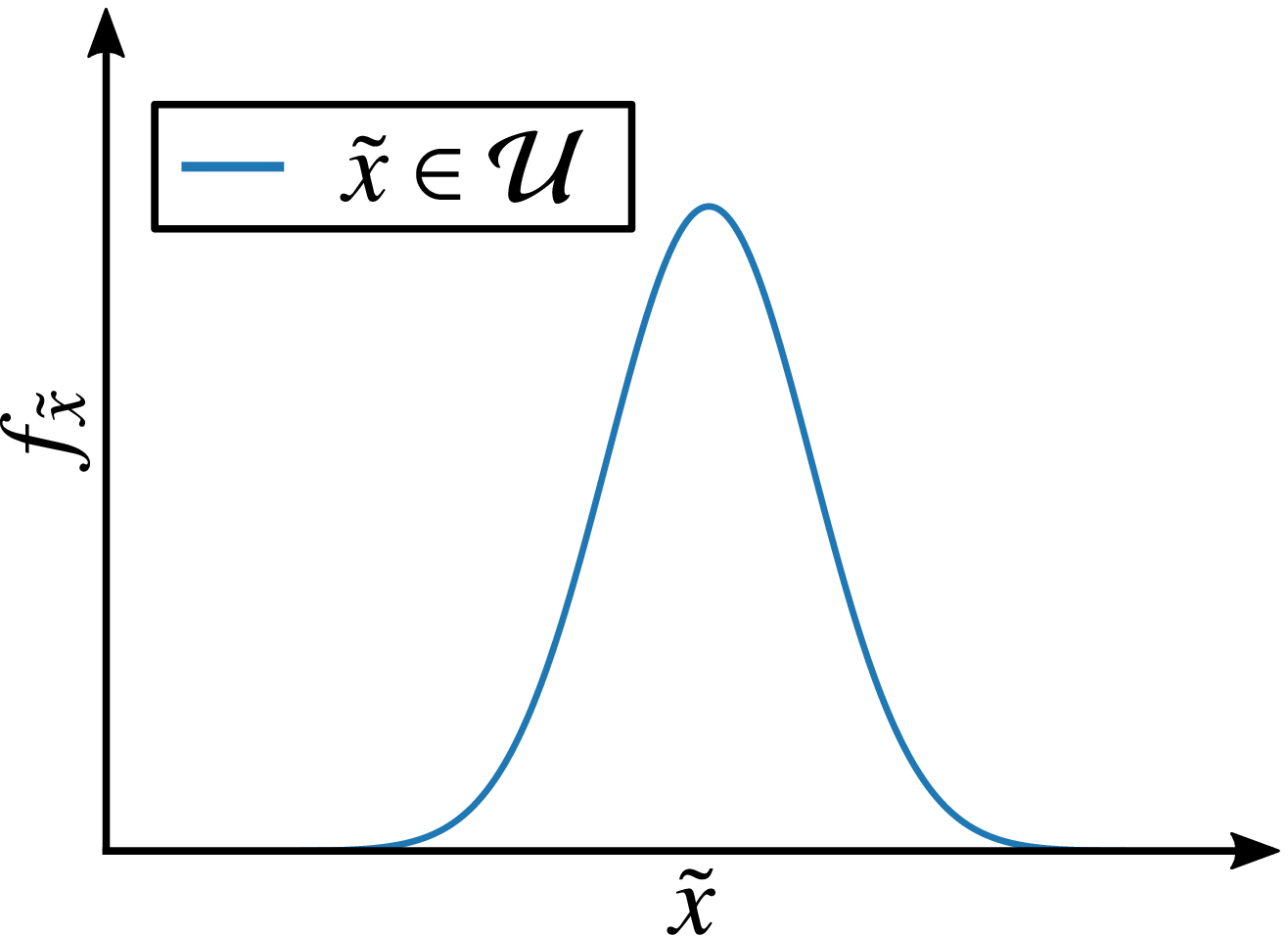}
\caption{Time-independent uncertain.}
\label{subfig:Vu}
\end{subfigure}%
\hspace{0.01\columnwidth}%
\begin{subfigure}[t]{0.495\columnwidth}
\centering
\includegraphics[scale=0.32]{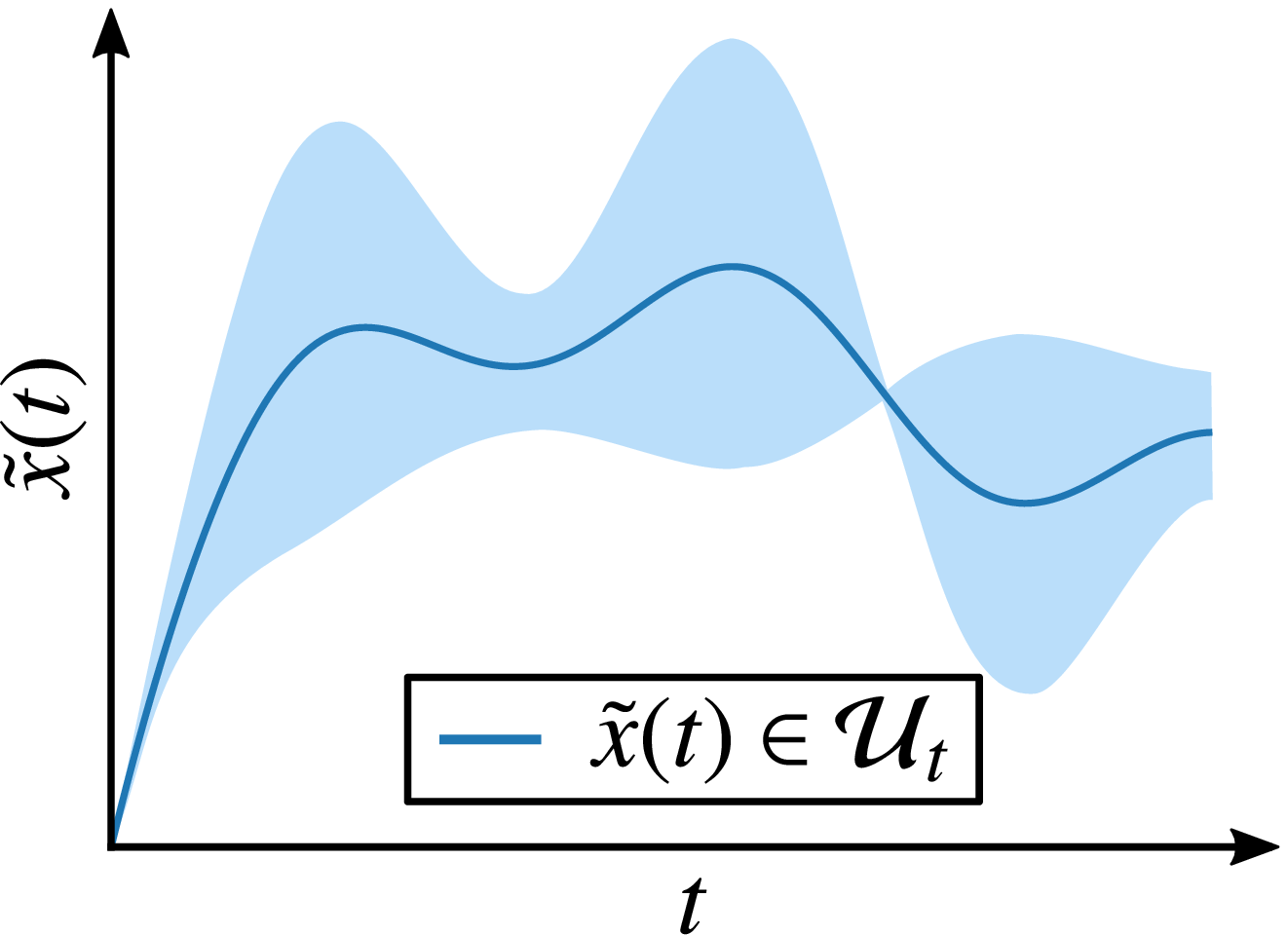}
\caption{Time-dependent uncertain.}
\label{subfig:Tu}
\end{subfigure}
\captionsetup[figure]{justification=centering}
\caption{Illustration of sets associated with time-independent and time-dependent deterministic and uncertain variables.}
\label{fig:foursets}
\end{figure}

% new paragraph
To better distinguish between these quantities in the future sections, we first define four general types of variables along with their associated sets.
Any arbitrary, time-independent deterministic variable $x$ is defined in the set $\mathcal{D}$.
As an example, $\mathcal{D}$ may be the set of real numbers $\mathbb{R}$, or natural numbers $\mathbb{N}$, or integers $\mathbb{Z}$, etc.
Figure~\ref{subfig:Vd} shows an arbitrary value belonging to $\mathcal{D}$. 

% Figure 1

% new paragraph
The set associated with a time-dependent deterministic variable (i.e.,~a trajectory) is defined as:    
\begin{align}
\mathcal{D}_{t} \coloneqq  \{ x(t) \mid t \in [t_{0}, t_{f}],~x(t) &\in \mathcal{D} \} 
\end{align}
\noindent
In other words, at every point in time, the trajectory is defined within a deterministic set. 
Figure \ref{subfig:Td} shows an arbitrary value belonging to $\mathcal{D}_{t}$. 

% new paragraph
For an arbitrary uncertain variable $\tilde{x}$, the sampling space is defined as $\mathcal{U}$.
As an example, $\mathcal{U}$ may be a set of time-independent uncertain variables with a Gaussian distribution, as shown in 
Fig.~\ref{subfig:Vu}.
Note that while the term sampling space implies uncertainty, it does not have any implications on probability.
Therefore, $\mathcal{U}$ may be an uncertainty set with or without a probability measure.

\begin{figure}[t]
\centering
\definecolor{myblue}{RGB}{31, 119, 180}
\definecolor{myred}{RGB}{214, 39, 40}
\tikzstyle{every node}=[draw=black, very thick, anchor=west, align=center,rounded corners=0.1cm]
\tikzstyle{selected}=[draw=myblue, fill=myblue!10, minimum width=2.2cm]
\tikzstyle{optional}=[draw=myred, fill=myred!10, text width=4.8cm]
\tikzstyle{methods2}=[draw=myblue, fill=myblue!10, text width=2.8cm, minimum width=2.8cm]
\tikzstyle{formulations2}=[draw=myred, fill=myred!10, text width=2.8cm]
\resizebox{0.85\columnwidth}{!}{%
\begin{tikzpicture}[%
  grow via three points={one child at (0.5,-1.1) and
  two children at (0.5,-1.1) and (0.5,-2.2)},
  edge from parent path={(\tikzparentnode.south) |- (\tikzchildnode.west)}, very thick]
  \node {\Large UCCD}
    child { node [selected] {Probabilistic}
        child { node [optional] {Stochastic Expectation (SE-UCCD) in Eq.~(\ref{Eqn:SCCDE})}}
        child { node [optional] {Stochastic Chance-Constrained (SCC-UCCD) in Eq.~(\ref{Eqn:CCCCD})}}
        child { node [optional] {Probabilistic Robust \\(PR-UCCD) in Eqs.~(\ref{Eqn:RCCD_p1})--(\ref{Eqn:RCCD_p2})}}
    }
    child [missing] {}				
    child [missing] {}				
    child [missing] {}
    child { node [selected] {Crisp}
      child { node [optional] {{Worst-Case} (Minimax) Robust (WCR-UCCD) in Eqs.~(\ref{Eqn:RCCDc})--(\ref{Eqn:RCCDd})}}
      }
    child [missing] {}
    child { node [selected] {Possibilistic}
      child { node [optional] {Fuzzy Expectation\\ (FE-UCCD) in Eq.~(\ref{Eqn:PCCCCDa})}}
      child { node [optional] {Possibilistic Chance-Constrained (PCC-UCCD) in Eq.~(\ref{Eqn:PCCCCD})}}
    }
    child [missing] {}
    child [missing] {};
    \node [draw=none] at (-1.6,-8) {Key:};
    \node [methods2] at (-1.6,-8.8) {Uncertainty Representation};
    \node [formulations2] at (-1.6,-10) {Specialized UCCD Formulations};
\end{tikzpicture}
}
\caption{Specialized UCCD formulations based on the uncertainty representation.}
\label{fig:UCCDclass}
\end{figure}
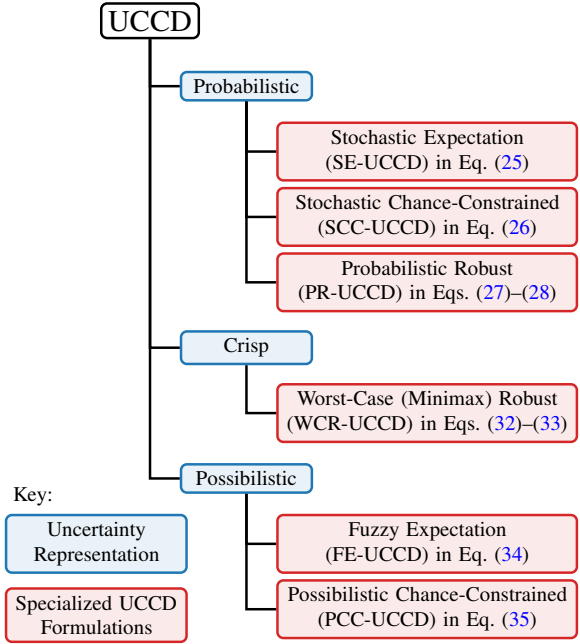

% new paragraph
Finally, for an arbitrary uncertain trajectory, the sampling space is defined as:
\begin{align}
\mathcal{U}_{t} \coloneqq  \{ \tilde{x}(t) \mid t \in [t_{0}, t_{f}],~\tilde{x}(t) \in \mathcal{U} \} 
\end{align}
\noindent
Similarly, $\mathcal{U}_{t}$ makes no assumptions regarding the probability measure.
Figure~\ref{subfig:Tu} shows an arbitrary time-dependent uncertain trajectory along with its associated sampling space.
Any uncertain variable belonging to $\mathcal{U}$ and $\mathcal{U}_{t}$ may be represented in three ways: (i) probabilistic, (ii) crisp, and (iii) possibilistic \cite{beyer2007robust}. 
In this article, we use these uncertainty representations to develop specialized UCCD formulations outlined in Fig.~\ref{fig:UCCDclass}.

% figure 2 Tikz with links

% Figure 3
\begin{figure}[t]
\centering
\begin{subfigure}[t]{0.495\columnwidth}
\centering
\includegraphics[scale=0.325]{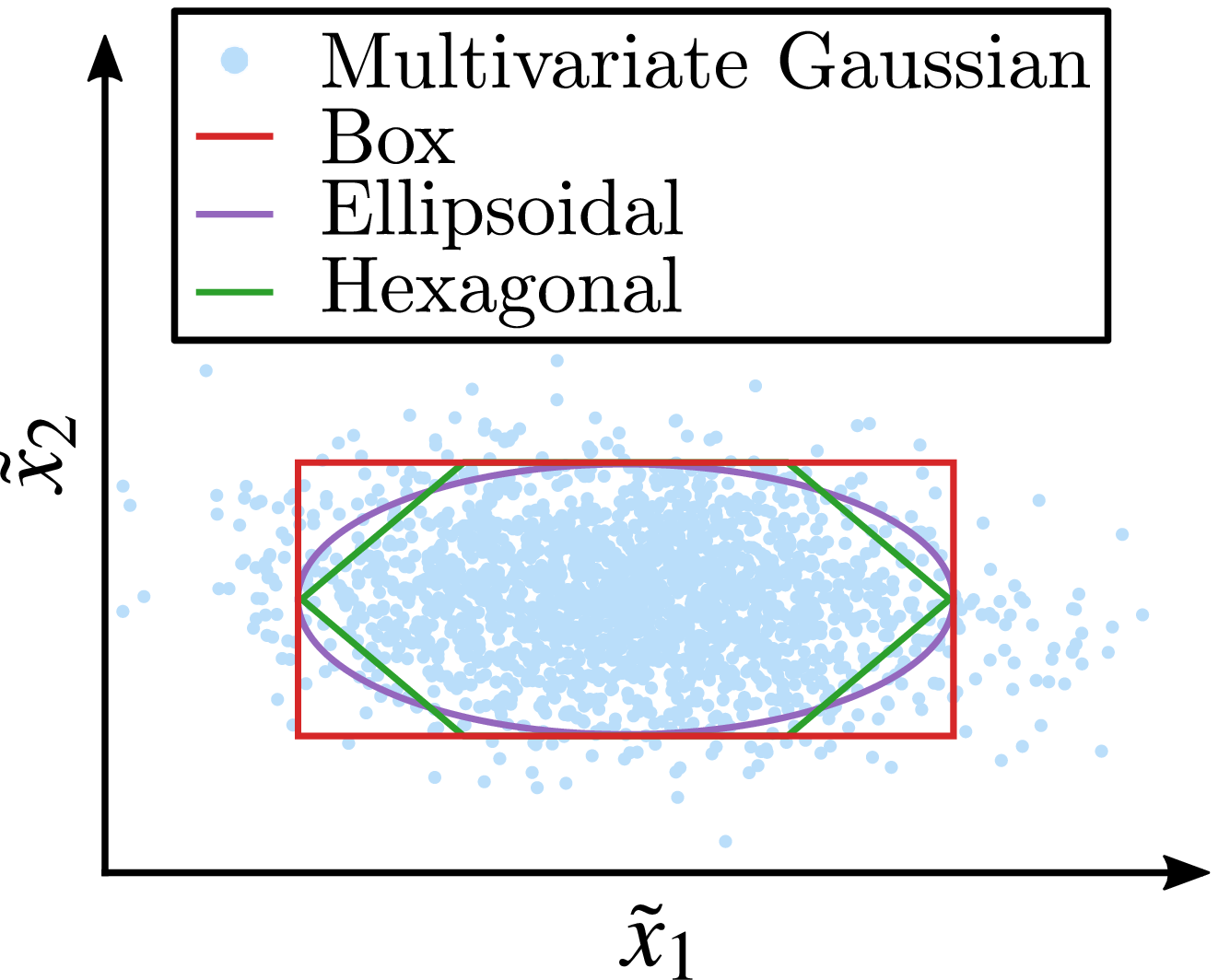}
\caption{}
\label{subfig:1}
\end{subfigure}%
\hspace{0.01\columnwidth}%
\begin{subfigure}[t]{0.495\columnwidth}
\centering
\includegraphics[scale=0.325]{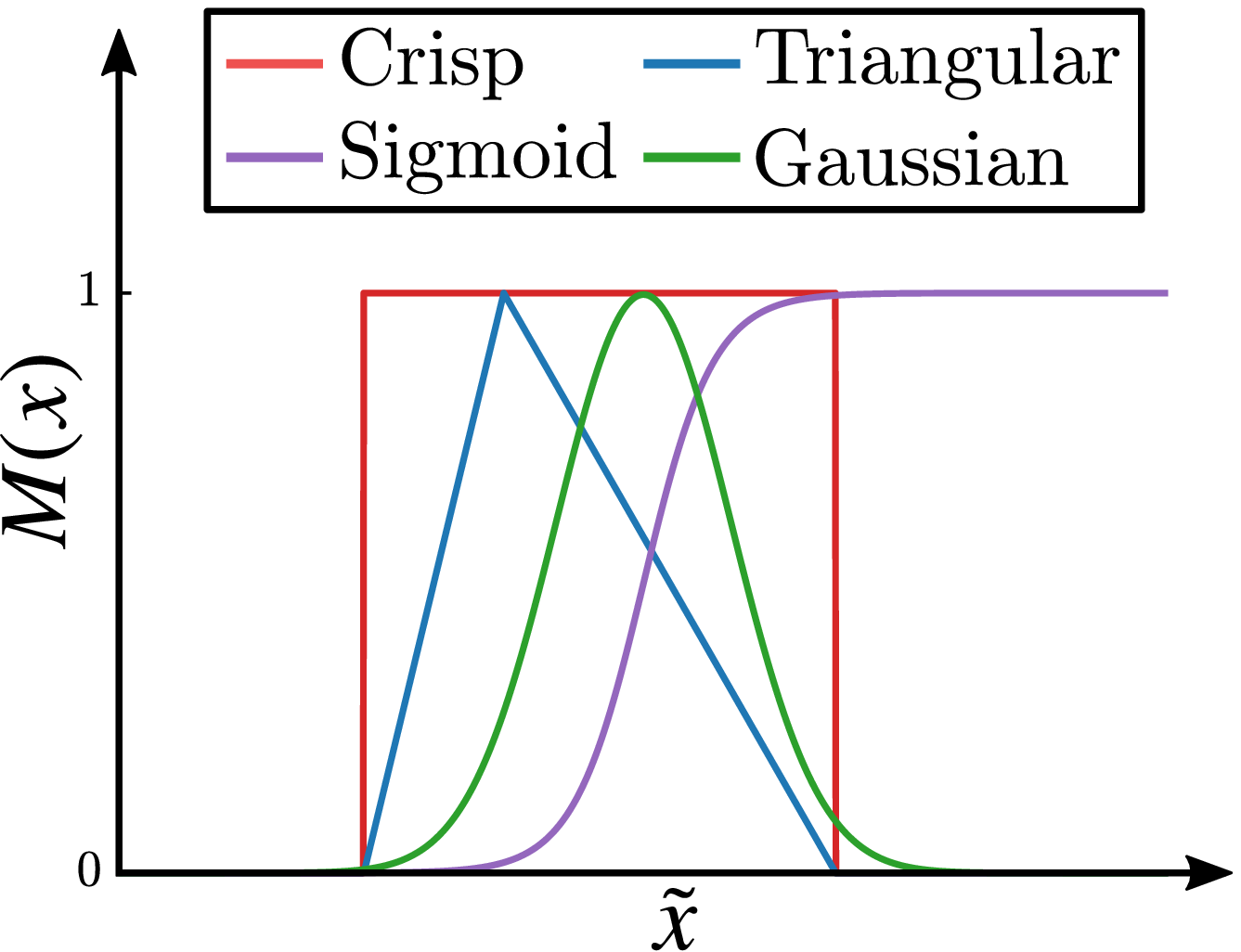}
\caption{}
\label{subfig:2}
\end{subfigure}%
\hspace{0.015\textwidth}
\caption{Various representations of uncertainties: (a) comparison of samples of Gaussian multivariate distribution for random variables $\tilde{x}_{1}$ and $\tilde{x}_{2}$ to the box, ellipsoidal, and hexagonal uncertainty sets and, (b)~several examples of fuzzy set membership functions.}
\label{fig:sets}
\end{figure}

\paragraph{Stochastic (Probabilistic).\label{par:PrR}}%
In the stochastic representation of uncertainties (also known as probabilistic), it is assumed that the associated probability distribution is known or can be estimated.
Therefore, if $\mathcal{U}$ and/or $\mathcal{U}_{t}$ is endowed with a probability measure, uncertainties can be described probabilistically.
For an arbitrary, time-independent, continuous uncertain variable $\tilde{x}$, the stochastic set is defined as: 
\begin{align}
    \label{StcSeta} \mathcal{X}_{\text{stc}} \coloneqq  \{ (\tilde{x}, F_{\tilde{x}}(x)) \mid \tilde{x} \in \mathcal{U},~F_{\tilde{x}}(x) = \mathbb{P}[\tilde{x} \leq x] \in [0, 1]  \}
\end{align}
\noindent
where the subscript $\text{stc}$ stands for stochastic, $\mathcal{X}_\text{{stc}}$ is the probabilistic set characterized by $F_{\tilde{x}}(x)$, which is the distribution function of $\tilde{x}$, and $x$ is a realization.
The probabilistic set for a time-dependent uncertain variable $\tilde{x}(t)$ is described as:
\begin{align}
    \label{StcSetb}    
    \mathcal{X}_\text{{stc}}(t) \coloneqq  \{ \tilde{x}(t) \mid t \in [t_{0}, t_{f}],~\tilde{x}(t) \in \mathcal{X}_\text{{stc}}  \}
\end{align}
\noindent
An example of the probabilistic representation of uncertainties is the assumption of a Gaussian distribution for uncertainties in a plant optimization variable's value.
Samples of a multivariate Gaussian distribution are shown in Fig.~\ref{subfig:1}.
This description of uncertainties motivates stochastic UCCD formulations.

\paragraph{\text{Crisp}.\label{par:DeR}}%
In the crisp representation of uncertainties, no probability measure is available, and uncertainties are assumed to belong to a crisp, deterministic set that can be finite, infinite, bounded, unbounded, discrete, or continuous.
For an arbitrary, time-independent uncertain variable $\tilde{x}$, the crisp representation of uncertainties entails a membership function that assigns one to all members and zero to all non-members: 
\begin{align}
    \label{CrispSeta}
    \mathcal{X}_{\text{crisp}} \coloneqq  \{ (\tilde{x},M_{\text{crisp}}(x)) \mid \tilde{x} \in \mathcal{U},~M_{\text{crisp}}(x) \in \{0,1 \}  \}\end{align}
\noindent
where $\mathcal{X}_{\text{crisp}}$ is the crisp set characterized by its associated membership function $M_{\text{crisp}}(x)$.
For a time-dependent uncertain variable $\tilde{x}(t)$, the crisp representation is described as:
\begin{align}
    \label{CrispSetb}
    \mathcal{X}_{\text{crisp}}(t) \coloneqq  \{ \tilde{x}(t) \mid t \in [t_{0}, t_{f}],~\tilde{x}(t) \in \mathcal{X}_{\text{crisp}}  \}
\end{align}
\noindent
Figure~\ref{subfig:1} compares samples from an arbitrary multivariate Gaussian distribution to the bounded, crisp representation of uncertainties associated with box, ellipsoidal, and hexagonal sets. 
Among these uncertainty sets, the box and hexagonal uncertainty sets are convex polytopes.
For linear programs, when uncertainties are restricted to a polytope, the number of function evaluations for uncertainty propagation may be reduced to function evaluations at the vertices of the polytope \cite{Azad2022b}.

\paragraph{Possibilistic.\label{par:PoR}}%
Uncertainty representations discussed so far are based on some available information, i.e.,~the known (or estimated) probability distribution function or geometry and size of the uncertainty set.
However, when too little is known about the uncertainty, one might utilize descriptive (and often vague) language (also known as linguistic variables) to express the desired or expected events.
This information is interpreted by an expert in the field and is best represented through a fuzzy set, which is a class with a continuum of grades of membership.

% new paragraph
For an arbitrary, time-independent uncertain variable $\tilde{x}$, the fuzzy set is defined as:
\begin{align}
    \label{FuzzySeta}
    \mathcal{X}_{\text{fuzzy}} \coloneqq  \{ (\tilde{x},M_{\text{fuzzy}}(x) \mid \tilde{x} \in \mathcal{U},~M_{\text{fuzzy}}(x) \in [0,1]  \}
\end{align}
\noindent
where $\mathcal{X}_{\text{fuzzy}}$ is the fuzzy set characterized by its associated membership function $M_{\text{fuzzy}}(x)$.  
This membership function practically quantifies the degree of membership of an element, or the possibility that an element belongs to the set\textemdash leading to concepts from possibility theory \cite{zadeh1996fuzzy, zadeh1978fuzzy}. 
For a time-dependent variable $\tilde{x}(t)$, the fuzzy set is defined as:
\begin{align}
    \label{FuzzySet}
    \mathcal{X}_{\text{fuzzy}}(t) \coloneqq  \{ \tilde{x} \mid t \in [t_{0}, t_{f}],~\tilde{x}(t) \in \mathcal{X}_{\text{fuzzy}}  \}
\end{align}
\noindent
Figure~\ref{subfig:2} compares the membership function of a crisp uncertainty set to that of a triangular, sigmoid, and Gaussian fuzzy membership functions.

\subsection{Other Considerations}
In general, it is natural to assume that in an arbitrary UCCD problem, uncertainties are represented based on the availability of information.
The choice of uncertainty representation, to some degree, informs the associated class of formulation.
Despite that, the decision-making process may entail other factors that ultimately demand an alternative choice of uncertainty representation.
For instance, the risk associated with specific performance criteria may be so critical that no constraint violation can be tolerated.
In this case, even if the distributional information is available, a worst-case robust formulation (see Secs.~\ref{par:WoD} and \ref{subsec:wcRCCD}) may be more practical. 

% new paragraph
A general UCCD problem may entail known uncertainties requiring one or more of the aforementioned representations.
Therefore, comprehensive treatment of uncertainties in UCCD problems requires the development of hybrid methods that are adept at integrating, combining, and interpreting all of such known uncertainties. 
These methods are generally referred to as hybrid programming \cite{liu2009theory} and have not yet been investigated for UCCD problems. 
It is also important to note that many real-world systems may also entail some unknown unknowns.
These are uncertainties that we don't know we don't know.  
Unknown unknowns will most likely be present in UCCD formulations and require additional protective measures \cite{zio2013literature}. 
In this article, we only focus on known unknowns.

%--------------Section 3-----------------%
\xsection{Mathematical Foundations for UCCD Formulations}
\label{sec:3}
In this section, we start by introducing a generalized UCCD problem formulation using concepts from probability theory.
Defining this formulation in the probability space is without any loss of generality because specialized forms of this formulation can be derived through the appropriate selection of the objective function and constraints.
This is specifically evident for crisp uncertainty sets as the associated expectation of the objective function and constraints reduce to deterministic quantities.
For fuzzy uncertainties, several formulations become viable, such as deterministic (crisp) formulation \cite{filev1992fuzzy}, expected value \cite{liu2002expected, zhu2009fuzzy}, optimistic/pessimistic, and credibility measures \cite{liu2009theory}. 
Due to the general correlation between operators in the probability and fuzzy space, specialized problem formulations in the fuzzy space can also be derived from the proposed formulation.
The generalized UCCD formulation is capable of capturing uncertainty descriptions that are introduced in this article.
Such descriptions are commonly used in areas such as control co-design, optimal control, operations research, robust design, and reliability-based design optimization and encompass a large portion of uncertainty-based considerations in the literature.

\paragraph{Preliminaries.}%
The stochastic modeling of any arbitrary vector $\bm{x} \in \mathbb{R}^{n_{x}}$ consists in introducing a sampling space $\Theta$ (such that any element of $\Theta$ is a combination of causes that affect the state of $\bm{x}$), and then endowing it with an event space $\mathcal{F}$, and a probability measure $\mathbb{P}$, which results in the probability space $(\Theta, \mathcal{F}, \mathbb{P})$ \cite{soize2017uncertainty}.
A stochastic variable $\tilde{\bm{x}} = (\tilde{x}_{1}, \dots, \tilde{x}_{n_{x}})$ defined on $(\Theta, \mathcal{F}, \mathbb{P})$ and endowed with a measurable space is then a mapping from $\Theta$ to $\mathbb{R}^{n_{x}}$ such that $\tilde{\bm{x}} \in \mathcal{X}_{\text{stc}}$.
A stochastic process $\tilde{\bm{x}}(t) \in \mathcal{X}_{\text{stc}}(t)$, is defined on the probability space and has values in $\mathbb{R}^{n_{x}}$.
$\tilde{\bm{x}}(t)$ is indexed by any finite or infinite subset $T$ and is a mapping $t \mapsto \tilde{\bm{x}}(t)$ from $T \times (\Theta, \mathcal{F}, \mathbb{P})$ into $L^{0}(\Theta, \mathbb{R}^{n_{x}})$.
Here, $L^{0}(\Theta, \mathbb{R}^{n_{x}})$ is the vector space of all $\mathbb{R}^{n_{x}}$-valued random variables defined on $(\Theta, \mathcal{F}, \mathbb{P})$.
For any fixed $\theta \in \Theta$, the mapping $t \mapsto \tilde{\bm{x}}(t,\theta)$ is a trajectory or a sample path. 
For an arbitrary stochastic variable $\tilde{\bm{x}}$, $\bm{x}_{\mu}$ is the mean value and $\bm{x}_{\sigma}$ is the standard deviation. 
In addition, $\mathbb{P}[\cdot]$ is the probability measure, and $\mathbb{E}[\cdot]$ is the expected value operator. 
For an arbitrary function of random variables, ${o}(\tilde{\bm{x}})$, its expected value is defined as $\mathbb{E}[{o}(\tilde{\bm{x}})] = \int_{-\infty}^{\infty} \cdots \int_{-\infty}^{\infty} {o}(\bm{x})f_{\tilde{\bm{x}}}(\bm{x}) dx_{1} \cdots dx_{n_{x}}$ for a continuous random vector and $\mathbb{E}[{o}(\tilde{\bm{x}})] =  \sum_{{x}_{1}} \cdots \sum_{{x}_{n_{x}}} {o}(\bm{x})p_{\tilde{\bm{x}}}(\bm{x})$ for a discrete random vector.
In these definitions, $f_{\tilde{\bm{x}}}(\bm{x})$ and $p_{\tilde{\bm{x}}}(\bm{x})$ are the probability distribution functions and mass functions, respectively.
We use $\bar{\parm}(\cdot)$ to indicate a specified function composition of $\parm(\cdot)$.
Specifically, $\bar{o}(\cdot)$ describes a function of the original objective function $o(\cdot)$ such that $\bar{o} = y \circ o(\cdot) = y(o(\cdot))$, where $y(\cdot)$ is an explicit or implicit function such that, when uncertainties are not present, $\bar{o}(\cdot)$ is reduced to its original deterministic form $o(\cdot)$.
With these definitions, the generalized UCCD problem formulation can be introduced.

\subsection{A Generalized UCCD Formulation}
\label{subsec:UCCCformulation}
A generalized, AAO, continuous-time, simultaneous UCCD problem can be formulated as:
\begin{subequations}
 \label{Eqn:UCCD}
 \begin{align}
 \underset{\tilde{\bm{u}}, \tilde{\bm{\xi}}, \tilde{\bm{p}}}{\textrm{minimize:}}
 \quad & \mathbb{E}\left [\bar{o} ( t, \tilde{\bm{u}}, \tilde{\bm{\xi}}, \tilde{\bm{p}}, \tilde{\bm{d}}) \right]  \label{UCCD_obj} \\
 \textrm{subject to:} \quad
&  \mathbb{E}\left [ \bar{\bm{g}} ( t, \tilde{\bm{u}}, \tilde{\bm{\xi}}, \tilde{\bm{p}}, \tilde{\bm{d}}) \right] \leq {\bm{0}} \label{UCCD_ineq} \\
& {\bm{h}}(t, \tilde{\bm{u}}, \tilde{\bm{\xi}}, \tilde{\bm{p}}, \tilde{\bm{d}})={\bm{0}}  \label{UCCD_eq}  \\
& \dot{\tilde{{\bm{\xi}}}}(t) - {\bm{f}}(t, \tilde{\bm{u}}, \tilde{\bm{\xi}}, \tilde{\bm{p}}, \tilde{\bm{d}}) = \bm{0} \label{UCCD_dyn} \\
\textrm{where:} \quad & \tilde{\bm{u}} = \tilde{\bm{u}}(t) \in \mathcal{U}_{t},~\tilde{\bm{\xi}} = \tilde{\bm{\xi}}(t) \in \mathcal{U}_{t} \label{UCCD_where}\\
& \tilde{\bm{p}} \in \mathcal{U},~\tilde{\bm{d}} = \tilde{\bm{d}}(t) \in \mathcal{U}_{t} \notag %\label{UCCD_td} 
 \end{align} 
\end{subequations}
\noindent
In this equation, the expectation of the composite function $\bar{o}(\cdot)$ (i.e.,~a function of the original objective $o(\cdot)$) is optimized over the set of optimization variables $(\tilde{\bm{u}}, \tilde{\bm{\xi}}, \tilde{\bm{p}})$, and is subject to the expectation of the composite functions $\bar{\bm{g}}(\cdot)$ (i.e.,~functions of the original inequality constraint vector $\bm{g}(\cdot)$), analysis-type equality constraints ${\bm{h}}(\cdot)$, and uncertain dynamic system equality constraints in Eq.~(\ref{UCCD_dyn}). 
Note that $\mathbb{E}[\bar{o}(\cdot)]$ and $\mathbb{E}[\bar{\bm{g}}(\cdot)]$ refer to any of the variations that will be discussed in Sec.~\ref{subsec:IC} (such as the nominal, worst-case, expected value, etc.).

% new paragraph
This formulation includes the vector of uncertain control processes
$\tilde{\bm{u}}(t) \in  \mathcal{U}_{t}$, uncertain state processes $\tilde{\bm{\xi}}(t) \in \mathcal{U}_{t}$, time-independent uncertain optimization variables $\tilde{\bm{p}} \in \mathcal{U}$, and
time-dependent $\tilde{\bm{d}}(t) \in \mathcal{U}_{t}$ and/or time-independent uncertain problem data $\tilde{\bm{d}} \in \mathcal{U}$.
Note that $\tilde{\bm{d}}(t)$ may entail some noise or disturbances that affect system dynamics.
Such uncertainties generally enter through the dynamic system model and captured through Eq.~(\ref{UCCD_dyn}) (see Refs.~\cite{nash2021robust,yang2011robust}). 

% new paragraph
The proposed UCCD formulation is infinite-dimensional in time and uncertainty dimensions. 
We can draw an analogy between the infinite-dimensional time vector and the infinite-dimensional uncertainty vector. 
To transcribe Eq.~(\ref{Eqn:UCCD}) in time, numerical methods such as direct transcription have been implemented~\cite{allison2014special,allison2014co,betts2010practical,biegler2010nonlinear,rao2009survey}.
Similarly, different uncertainty propagation techniques, such as Monte Carlo simulation (MCS), generalized polynomial chaos, as well as special interpretations, such as worst-case, have been proposed to parameterize the uncertain dimensions \cite{Azad2022b}.
In this article, we discuss some of these formulations and special considerations but generally leave the discussion on specialized solution methods to future work.
By emphasizing various uncertainty interpretations through generalized and then specialized formulations, this article aims to provide an improved understanding of some of the design challenges and insights in the presence of uncertainties.

% new paragraph
We emphasize that describing optimization variables $(\tilde{\bm{u}}, \tilde{\bm{\xi}}, \tilde{\bm{p}})$ in the uncertain space is to avoid introducing any unnecessary assumptions/structure at this point.
Furthermore, this description should not imply that the designer has complete control over all uncertainties; instead, it suggests that the decision space may entail elements associated with uncertainties.
In other words, these uncertain quantities may entail some deterministic part the designer/optimizer has decision-making power over.
This deterministic part may be associated with the mean values, parameters of an entire distribution, shape or geometry of the deterministic uncertainty set, or parameters of the fuzzy membership function.

\subsection{Uncertainties in Optimization Variables}
\label{subsec:SoU}
CCD is an enticing approach because it simultaneously explores the plant and control design spaces to improve the dynamic system's performance \cite{allison2014special}.
When uncertainties are present, it is imperative to maintain this advantage by introducing balanced UCCD formulations in which the whole space of optimization variables is leveraged in response to uncertainties. 
Therefore, in a balanced formulation, uncertain control and state trajectories, as well as the vector of time-independent optimization variables, must be utilized to achieve a system-level, integrated solution. 
To accomplish this vision, it is critical to understand where uncertainties in optimization variables $(\tilde{\bm{u}}, \tilde{\bm{\xi}}, \tilde{\bm{p}})$ originate from and how they affect various elements of the problem.

\subsubsection{Control Trajectories.}\label{par:CT}~In Eq.~(\ref{Eqn:UCCD}), control trajectories are modeled as stochastic processes because $\tilde{\bm{d}}$ may entail noise elements (induced by factors such as electrical noise, actuator imprecision, etc.) that directly affect control signals.
This, in the control community, is referred to as matching (or lumped) uncertainties because uncertainties act on the system through the same channels as the control input.
If uncertainties do not act through the control channel, they are called mismatched uncertainties \cite{yang2011robust}. 
Therefore, the above formulation entails both matched and mismatched uncertainties.
However, it is possible to model the control input deterministically since possible disturbances on the control can be modeled in the dynamics as multiplicative noise \cite{greco2020direct}.
Note that ``closing the control loop'' with feedback controller architectures in a UCCD problem may also transform the control trajectories into stochastic quantities.
Reference~\cite{lavretsky2007stable} describes the development and application of a reference adaptive control design scheme with matched uncertainties for an F-16 aircraft case study.%

\subsubsection{State Trajectories.}\label{par:ST}~In Eq.~(\ref{Eqn:UCCD}), state trajectories are uncertain due to a variety of reasons.  The uncertainties from ($\tilde{\bm{u}}, \tilde{\bm{p}}, \tilde{\bm{d}}$) may propagate through the dynamic system and transform them into stochastic processes.
Note that the resulting stochastic systems are not necessarily the same as the classical stochastic differential equations where the inputs are some idealized processes, such as Wiener or Poisson~\cite{xiu2010numerical}.
The vector of problem data, $\tilde{\bm{d}}$, may entail some information about uncertain initial/final conditions.
In addition, $\tilde{\bm{d}}$ entails some noise elements that may enter the state equation in a linear or nonlinear manner. 
This noise may be stationary or non-stationary, exogenous (independent of decisions), or dependent on states and controls.
As an example of the dependence of noise on states and controls, consider a system that starts to witness more chaotic changes after it is steered through the control command to a specific state.
However, note that this dependency is already captured through the dynamic model in Eq.~(\ref{UCCD_dyn}).
Note that $\tilde{\bm{\xi}}$ may also entail variables that are being controlled, or parameters of a distribution (such as mean and variance) describing the time-evolution of uncertainties in the system.
However, the distributional (or set) information of these parameters is specified and already included in the vector of uncertain problem data $\tilde{\bm{d}}$.
Also, note that the effects of unmodeled, mismodeled, and neglected dynamics can be captured in Eq.~(\ref{Eqn:UCCD})~\cite{yang2011robust}.
The implementation of a robust adaptive fuzzy tracking controller for a hypersonic flight vehicle subject to uncertainties from unmodeled and neglected dynamics is discussed in Ref.~\cite{hu2018robust}.

\subsubsection{Time-independent Optimization Variables.}\label{par:TI}~The vector of time-independent optimization variables may also be uncertain due to factors such as imperfect manufacturing processes, plant measurement errors, or mass production of plants.
In addition, over time, the dynamics of the plant may change (e.g., due to aging), which causes deviations compared to the original model. This deviation is known as model plant mismatch \cite{badwe2009detection}.
Therefore, $\tilde{\bm{p}}$ is modeled as a random variable whose distributional (or set) information is known.
This uncertainty will be propagated through state equations, transforming all of its associated parameter-dependent functions and variables into uncertain quantities.
In addition, for free-final-time UCCD problems, uncertainties may transform $t_{f}$ into an uncertain variable, requiring a transformation similar to the one described in Ref.~\cite{herber2017unified}.
Reference~\cite{azad2021robust} investigates the impact of time-independent uncertainties on the CCD solution of a hybrid-electric vehicle powertrain.

\subsection{Risk in UCCD Formulations}
\label{subsec:R}
In a UCCD formulation, uncertainties must be represented in a way that their impact on decision-making is completely captured.
This brings us to the notion of risk, which is a fundamental element of any uncertain problem.
In general, risk measures can be qualitative or quantitative \cite{andrieu2007stochastic}.
In a qualitative risk measure, the amount by which a threshold is surpassed does not matter.
An example of qualitative risk measures are failures that result in the loss of life.
In quantitative risk measures, on the other hand, it is important to know the extent to which the threshold is violated.
For example, a quantitative risk measure may be associated with the energy consumption of a vehicle following a reference trajectory.
When the energy consumption exceeds the threshold, it is important to know by how much.
This type of risk measure can be dealt with by introducing a penalty term or constraining the amount of extra energy.
In general, due to mathematical difficulties associated with probabilistic constraints, it is recommended to use probabilistic descriptions only for qualitative failure problems.
Other risk measures, such as conditional value-at-risk that offer mathematical properties (such as convexity), may be more suitable for quantitative constraint problems \cite{andrieu2007stochastic,rockafellar2002conditional}.

% new paragraph
The notion of risk is so central in decision-making under uncertainty that it is used to classify various problem classes based on the designer's attitude toward risk.
These include risk-neutral, risk-averse, risk-aware, and risk-sensitive problem formulations.
It is the designer's understanding of the risk associated with uncertainties in an arbitrary problem that determines the associated risk attitude in that formulation.
The focus of this article is mainly on risk-neutral and risk-averse UCCD formulations.
References~\cite{baringo2020optimal} and \cite{nakka2022trajectory} present a risk-neutral and risk-averse approach for optimal scheduling of a virtual power plant and motion planning of a robotic system, respectively.

\subsection{Objective Function in Epigraph Form}
\label{subsec:OF}
While some of the elements of a UCCD problem require specific treatment in the presence of uncertainties, an important point to emphasize is that there's no conceptual distinction between the treatment of an objective function and inequality constraints \cite{rockafellar2007coherent}.
This statement is without any loss of generality because, for any uncertain UCCD problem, the uncertain objective function may be transferred to the vector of inequality constraints through the addition of a new decision variable. 
This form is referred to as the epigraph representation of the objective function and allows us to deal with all of the complications resulting from uncertainties separately within inequality constraints.
Depending on the problem structure and the extent to which uncertainties affect various elements of the formulation, one may decide to keep or transfer the objective function.
The computational efficiency and resulting implications of such decisions on various classes of UCCD problems remain to be investigated.
The treatment of an uncertain objective function as an inequality constraint using epigraph representation for a simple strain-actuated solar array system is demonstrated in Ref.~\cite{Azad2022b}.

\subsection{Inequality Constraints}
\label{subsec:IC}
The formulation presented in Eq.~(\ref{Eqn:UCCD}) allows us to select $\bar{o}(\cdot)$ and $\bar{\bm{g}}(\cdot)$ in order to formulate various desired forms of the objective function and constraints.
In this section, these formulations are described only for the uncertain vector of inequality constraints $\bm{g}(\cdot)$. 
However, the same principles can be applied to formulate the objective function per the discussion in Sec.~\ref{subsec:OF}.

\subsubsection{Nominal.\label{par:NoD}}~In this formulation, uncertain quantities are prescribed and evaluated at their nominal (deterministic) values. This concept, which is referred to as guessing the future~\cite{rockafellar2007coherent}, attempts to estimate the unknown information for uncertain quantities.
As an example, instead of creating a probabilistic model for wind velocity at a given altitude, one may use a fixed, nominal input to evaluate and solve the problem.
This estimate, however, does not capture the impacts of uncertainties and makes no practical provisions for the risk associated with such uncertainties.
Recalling that the expected value of a deterministic term is a deterministic quantity, Eq.~(\ref{UCCD_ineq}) can be formulated by selecting a nominal value for uncertain factors:
\begin{equation}
\label{Eqn:nominalobjIneq}
\begin{aligned}
\mathbb{E}\left[ {\bar{g}_{i}} ( t, \tilde{\bm{u}}, \tilde{\bm{\xi}}, \tilde{\bm{p}}, \tilde{\bm{d}}) \right] = g_{i}(t, \bm{u}_{N},\bm{\xi}_{N},\bm{p}_{N}, \bm{d}_{N})  \leq 0
\end{aligned}
\end{equation}
\noindent
where $\parm_{N}$ refers to the nominal values of uncertain quantities in the $i$th inequality constraint.
As as example, Ref.~\cite{allison2014co} employs a nominal rough road profile for CCD of an active suspension system.

\subsubsection{Expected Value.\label{par:ExD}}~One of the most common probabilistic descriptions of uncertain inequality constraints is to utilize their corresponding average values~\cite{greco2020direct, nakka2022trajectory,  powell2019unified, girardeau2010comparison}.
In the stochastic programming community, this formulation is known as the expected value model.
This description, however, does not hedge against the risks associated with constraint violation.
Therefore, the expected value model is more suitable for objective function descriptions or risk-neutral formulations.  
As an example, the expected value model may be used to maximize the average energy production of a wind farm.  
The expected value model for the $i$th constraint is described as:
\begin{equation}
\label{Eqn:exobjIneq}
\begin{aligned}
\mathbb{E}\left [ {\bar{g}}_{i} ( t, \tilde{\bm{u}}, \tilde{\bm{\xi}}, \tilde{\bm{p}}, \tilde{\bm{d}}) \right] = g_{\mu,i}(t, \tilde{\bm{u}}, \tilde{\bm{\xi}}, \tilde{\bm{p}}, \tilde{\bm{d}}) \leq 0
\end{aligned}
\end{equation}
\noindent
A risk neutral bidding model for wind power producers is presented in Ref.~\cite{alashery2018risk}.%

\subsubsection{Long-Run Expected Value.\label{par:LoD}}~The long-run expected average \cite{malikopoulos2015multiobjective,hernandez1996infinite}, which is also known as the infinite-horizon expected average, is important in applications where the horizon is considered infinite, and it is desired to minimize the cost per unit time or satisfy some constraints over this infinite horizon.
Similar to the expected value model, the long-run expected value is most suitable for the description of the objective function or risk-neutral formulations.
As an example, this model may be used to describe the objective of minimizing the long-run average cost in a stochastic manufacturing system \cite{sethi1998optimal}. 
While infinite-horizon problems may take different forms, here, we introduce the formulation with a discounted cost:
\begin{equation}
\label{Eqn:longrunexobjIneq}
\begin{aligned}
\mathbb{E}\left [ {\bar{g}_{i}} ( t, \tilde{\bm{u}}, \tilde{\bm{\xi}}, \tilde{\bm{p}}, \tilde{\bm{d}}) \right] = \limsup_{t\to\infty} \mathbb{E}\left [ g_{i} ( t, \tilde{\bm{u}}, \tilde{\bm{\xi}}, \tilde{\bm{p}}, \tilde{\bm{d}}, \gamma) \right] \leq 0 
\end{aligned}
\end{equation}
\noindent
where $\gamma \geq 0$ is a discount parameter and $\limsup$ is used to highlight that it is not known whether the limit exists. 
The discount rate is included to emphasize short-term rewards versus rewards that might be obtained in the distant future.
A long-run expected value implementation for online stochastic control of hybrid electric vehicles is discussed in Ref.~\cite{malikopoulos2015multiobjective2}.

\subsubsection{Higher-Order Moments.\label{par:HmD}}~Sometimes, the higher-order moments of an uncertain quantity, particularly its variance, are used as a measure to hedge against uncertainties.
This is motivated by the fact that expected value alone does not consider the distribution or worst-case characteristics of the outcome.
As an example, a risk measure might be defined to limit the standard deviation (or variance) of one of the performance criteria, such as ride comfort, in an automotive active suspension design.   
This can be accomplished by defining:
\begin{equation}
\label{Eqn:varobjIneq}
\begin{aligned}
{\mathbb{E}\left [ \bar{g}_{i} ( t, \tilde{\bm{u}}, \tilde{\bm{\xi}}, \tilde{\bm{p}}, \tilde{\bm{d}}) \right] =  \sqrt{\mathbb{E}[g_{i}(\cdot)^2] - g_{\mu,i}(\cdot)^2} = {g}_{i,\sigma} (\cdot) \leq \sigma_{a,i}} 
\end{aligned}
\end{equation}
\noindent
where $g_{i,\sigma}$ refers to the standard deviation of the constraint and $\sigma_{a,i}$ is the allowable standard deviation associated with the $i$th constraint.
This description, which is generally accompanied by the expectation or the nominal value of the constraint (or objective function) is studied in Refs.~\cite{azad2020robust, ruszczynski2003stochastic, nagy2004open, li2014aircraft, azad2021robust}, and is further discussed in Sec.~\ref{subsec:PR_robust}. 
An implementation of this type for aircraft robust trajectory optimization is presented in Ref.~\cite{li2014aircraft}.

\subsubsection{Conditional Value-at-Risk.\label{par:Cvar}}~
In addition to higher-order moments described in Sec.~\ref{par:HmD}, an alternative risk measure, known as conditional value-at-risk (CVaR), may be utilized \cite{andrieu2007stochastic, rockafellar2002conditional}. 
CVaR is the expected value of the worst scenarios (i.e.,~realizations).  
This risk measure leverages the distributional information of the quantity of interest to identify undesirable outcomes, thereby providing insights into decisions that reduce the risks involved with the perceived worst scenarios.
For the $i$th inequality constraint, CVaR is defined as:
\begin{equation}
\label{Eqn:CVaR}
\begin{aligned}
{\mathbb{E}\left [ \bar{g}_{i} ( t, \tilde{\bm{u}}, \tilde{\bm{\xi}}, \tilde{\bm{p}}, \tilde{\bm{d}}) \right] = \mathbb{E}[g_{i}(\cdot) \mid g_{i}(\cdot) \geq \alpha_{q}(\Gamma)] = g_{i,\text{CVaR}}(\cdot) }
\end{aligned}
\end{equation}

\noindent
where $\alpha_{q}(\Gamma)$ is the quantile function of the distribution of $g_{i}$ with $\Gamma$ being the confidence level, also known as value-at-risk CVaR.
Reference \cite{shi2018stochastic} develops a fault tolerant control strategy using CVaR for wind energy conversion systems.

\subsubsection{Expected Utility Theory.\label{par:EUT}}
~Normative decision theory, which is mainly concerned with how agents \textit{ought} to make decisions, typically utilizes some axioms to formalize the requirements associated with rational and logical decision-making. 
The decision-maker's preferences and risk attitude are often captured by selecting an appropriate \textit{utility function} $U(\cdot)$ that assigns a subjective value to each outcome.
In the presence of uncertainties, expected utility theory is a normative theory that attempts to find the action that results in maximum expected utility \cite{von2007theory}. 
The choice of the utility function is strongly dependent on decision-maker's preferences and risk attitude.
While a utility function is commonly used to represent an objective, here we use this representation for the $i$th constraint.
This is because, as mentioned in Sec.~\ref{subsec:OF}, the objective function may be transferred to the vector of inequality constraints through the epigraph representation.  
Utilizing the expected utility as a normative decision theory, the $i$th constraint is described as:
\begin{equation}
\label{Eqn:utilityIneq}
\begin{aligned}
{\mathbb{E}\left [ \bar{g}_{i} ( t, \tilde{\bm{u}}, \tilde{\bm{\xi}}, \tilde{\bm{p}}, \tilde{\bm{d}}) \right] = \mathbb{E}[U_{i}(t, \tilde{\bm{u}}, \tilde{\bm{\xi}}, \tilde{\bm{p}}, \tilde{\bm{d}})] = U_{\mu,i}(\cdot)  \leq 0} 
\end{aligned}
\end{equation}
\noindent
where, $U_{\mu,i}(\cdot)$ is the expected utility associated with the $i$th constraint.
An example of a utility function is discussed in Sec.~\ref{subsec:PR_robust}, and an application of expected utility theory for strategic route choice is presented in Ref.~\cite{razo2013rank}.

\subsubsection{Probabilistic Chance-Constrained.\label{par:PrcD}}~Sometimes, it is desirable to express and satisfy constraints in terms of the probability of an event.
For example, the probability that a constraint associated with stress or deflection on a part is satisfied within a given threshold.
This can be done by defining the $i$th constraint in terms of an indicator function of an arbitrary event $E$:
\begin{equation}
\label{Eqn:aux_probabilistic}
\begin{aligned}
    \mathbb{I}_{E}(t, {{\bm{u}}}, {\bm{\xi}}, {\bm{p}}, {\bm{d}}) = \begin{cases}
1~\textrm{if}~\{ {\bm{u}}, {\bm{\xi}}, {\bm{p}}, {\bm{d}}\} \in E\\
0~\textrm{if}~\{{\bm{u}}, {\bm{\xi}}, {\bm{p}}, {\bm{d}}\} \notin E\\
\end{cases} 
\end{aligned}
\end{equation}
\noindent
Then, the probability can be defined through the expectation of the indicator function:
\begin{equation}
\label{Eqn:obj_probabilistic1}
\begin{aligned}
\mathbb{E}\left [ {\bar{g}_{i}} ( t, \tilde{\bm{u}}, \tilde{\bm{\xi}}, \tilde{\bm{p}}, \tilde{\bm{d}}) \right]  = \mathbb{E}\left [ \mathbb{I}_{E}(t, {\bm{u}}, {\bm{\xi}}, {\bm{p}}, {\bm{d}})  \right] = \mathbb{P}[E]
\end{aligned}
\end{equation}
\noindent
This formulation is the basis for the well-known chance-constraint programs and has resulted in wide range of methods that attempt to handle uncertain constraints reliably by prescribing a target failure probability $\mathbb{P}_{f,i}$ for the $i$th constraint \cite{azad2020single, cui2020comparative, nakka2022trajectory}, such that: 
\begin{align}
\label{Eqn:probabilistic2}
\mathbb{P}[g_{i}(t, \tilde{\bm{u}}, \tilde{\bm{\xi}}, \tilde{\bm{p}}, \tilde{\bm{d}}) \geq 0] \leq \mathbb{P}_{f,i}
\end{align}
\noindent
An application of the probabilistic chance-constrained formulation to the trajectory optimization of robotic spacecraft simulator is presented in Ref.~\cite{nakka2022trajectory}.

% new paragraph
Alternative chance-constrained formulations can also be developed in which the emphasis is on the system performance.
For example, in a series configuration, the probabilistic system chance-constrained formulation is described as \cite{ba2006new}:
\begin{equation}
\label{Eqn:probabilistic3}
\begin{aligned}
\mathbb{P}_{sys} = \mathbb{P} \left [\bigcup_{i=i}^{n_g} g_{i}(t, \tilde{\bm{u}}, \tilde{\bm{\xi}}, \tilde{\bm{p}}, \tilde{\bm{d}}) \geq 0 \right ] \leq \mathbb{P}_{f,sys}
\end{aligned}
\end{equation}
\noindent
where $\mathbb{P}_{f,sys}$ is the system failure probability.
The system-level reliability for the design of an internal combustion engine case study is investigated in Ref.~\cite{nguyen2010single}.

\subsubsection{Worst-Case.\label{par:WoD}}~When uncertainties are represented as crisp sets, it is generally desired to solve the UCCD problem such that the resulting solution is feasible for all realizations of randomness within the specified uncertainty set.
This interpretation is equivalent to a worst-case design philosophy, in which every constraint is satisfied for its associated worst-case uncertainty realization within the uncertainty set.
As an example, consider the design of an automotive brake system subject to uncertainties from the road surface, velocity, temperature, etc. 
For such a design problem, it is imperative that the brake system is capable of bringing the vehicle to a halt within a reasonable amount of time under any circumstances.  
If the bounds on uncertainties are known, one can minimize the worst combination of uncertainties in order to make sure that the brake system performs well for all other cases.

% new paragraph
The parameters of the uncertainty set, which determine its characteristics such as shape, size, and geometry, are in fact a modeling choice.
In addition, these uncertainty sets are often defined using some nominal parameters.
For decision variables, the optimizer often has control over such parameters and uses them to navigate the design space.
For uncertain problem data, these nominal parameters are prescribed within the vector $\tilde{\bm{\bm{d}}}$.
These nominal parameters are formally described as $\hat{\bm{q}}^{T} = [\hat{\bm{p}}, \hat{\bm{\bm{d}}}]$ and $\hat{\bm{q}}_{t}^{T}(t) = [\hat{\bm{u}}(t), \hat{\bm{\xi}}(t), \hat{\bm{\bm{d}}}(t)]$, for time-independent and time-dependent problem elements, respectively. 
From here, we can define the time-independent uncertainty set as $\mathcal{R}(\hat{\bm{q}}) = \{\mathcal{R}(\hat{\bm{p}}) \times \mathcal{R}({\hat{\bm{d}}}) \} \subseteq \mathcal{X}_{\text{crisp}}$, and the time-dependent uncertainty set as $\mathcal{R}_{t}(\hat{\bm{q}}_{t}) =  \{ \mathcal{R}(\hat{\bm{u}}) \times  \mathcal{R}(\hat{\bm{\xi}}) \times \mathcal{R}(\hat{\bm{d}})     \} \subseteq \mathcal{X}_{\text{crisp}}(t)$.

 % new paragraph
In the worst-case description, the $i$th inequality constraint can then be represented as:
\begin{equation}
\label{Eqn:wcobjIneq}
\begin{aligned}
    \mathbb{E}\left [ {\bar{g}_{i}} ( t, \tilde{\bm{u}}, \tilde{\bm{\xi}}, \tilde{\bm{p}}, \tilde{\bm{d}}) \right] =
    \underset{\parbox{0.93in}{\centering {$(\bm{u}, \bm{\xi}, \bm{d}) \in \mathcal{R}_{t}(\hat{\bm{q}}_{t})$}\\ {$\bm{p} \in \mathcal{R}(\hat{\bm{q}})$}}}{\textrm{maximize}}
     \left\lbrace g_{i}(t, {\bm{u}}, {\bm{\xi}}, {\bm{p}}, {\bm{d}})  \right\rbrace \leq 0
\end{aligned}
\end{equation}

\noindent
When this treatment is applied for the complete optimization problem, it results in a bi-level formulation known as the $\min-\max$ or minimax\cite{nash2021robust,zhang2017robust}. 
We note that for UCCD problems, this maximization problem must be solved subject to analysis-type system equality constraints, which will be discussed in more detail in Sec.~\ref{subsec:wcRCCD}. 
This description is used to find a robust UCCD solution of an aircraft thermal management system using model predictive control in Ref.~\cite{nash2021robust}.

\subsubsection{Possibilistic Chance-Constrained.\label{par:PcD}}~When uncertainties are defined through fuzzy variables/processes, equivalent chance-constrained formulations may be developed in the possibility space.
As an example, when little information is known about uncertainties in the vehicle side-impact performance problem, one may formulate a chance constraint such that the possibility of failure is below a given threshold. 
The associated possibility-based constraint can be written as: 
\begin{equation}
    \label{Eqn:fuzzycons}
    \mathbb{P}\mathbb{O}\mathbb{S} [g_{i}(t, \tilde{\bm{u}}, \tilde{\bm{\xi}}, \tilde{\bm{p}}, \tilde{\bm{d}}) \geq 0] \leq \mathbb{P}\mathbb{O}\mathbb{S}_{f,i}
\end{equation}
\noindent
where $\mathbb{P}\mathbb{O}\mathbb{S}[\cdot]$ is the possibility measure defined on a proper possibility space, and $\mathbb{P}\mathbb{O}\mathbb{S}_{f,i}$ is the failure possibility for $i$th constraint.
For the sake of brevity, in this article, we avoid a detailed mathematical description of the possibility space and refer the readers to Refs.~\cite{liu2002expected,zhu2009fuzzy,liu2002toward} for further discussion.
A possibilistic framework for the design of unmanned electric vehicles is discussed in Ref.~\cite{shen2022possibilistic}.

\subsubsection{Dempster-Shafer (Evidence) Theory.\label{par:Evidence}}
Evidence theory, also known as the theory of belief measures,  deals with situations where limited information on uncertainties is available.
As opposed to probability theory which offers only a single measure (i.e.,~probability), evidence theory provides two uncertainty measures, known as belief and plausibility, both of which are determined from the known evidence for a proposition \cite{agarwal2004uncertainty}.
This evidence, also referred to as a body of evidence, is characterized by the basic probability assignment (BPA) function.  
Belief and plausibility give the lower and upper bounds of the event, respectively. 
\begin{equation}
\label{Eqn:evidence}
\begin{aligned}
{\mathbb{E}\left [ \bar{g}_{i} ( t, \tilde{\bm{u}}, \tilde{\bm{\xi}}, \tilde{\bm{p}}, \tilde{\bm{d}}) \right] = U_{i,r}^{M} - U_{i}^{M}(t, \tilde{\bm{u}}, \tilde{\bm{\xi}}, \tilde{\bm{p}}, \tilde{\bm{d}})  \leq 0} 
\end{aligned}
\end{equation}
\noindent
where $U_{i}^{M}$ is either belief or plausibility and $U_{i,r}^{M}$ is the required uncertainty measure for $i$th constraint.
Interested readers may refer to Refs.~\cite{agarwal2004uncertainty,bae2004approximation} for further details.

% new paragraph
The formulations introduced above are among the common descriptions of uncertain inequality constraints (and objective functions). 
Other variations exist that generally attempt to address some of the shortcomings of these formulations. 
For example, multiple formulations, such as min-max regret models, have been developed to address the issue of the conservativeness of the minimax approach \cite{schobel2014generalized}.

\subsection{Equality Constraints}
\label{subsec:EC}
In the presence of uncertainties, equality constraints are divided into two categories \cite{mattson2003handling, rangavajhala2007challenge}:~(i) those that must be strictly satisfied regardless of uncertainties (Type~I), and (ii) those that cannot be strictly satisfied due to uncertainties (Type~II).

% new paragraph
Type~I equality constraints, which are also referred to as analysis-type constraints, generally describe the laws of nature or dynamics of the system, such as Eqs.~(\ref{UCCD_eq}) and (\ref{UCCD_dyn}).
Therefore, for the problem to be physically meaningful, these constraints must be strictly satisfied at all parameterized points along the uncertain dimension.
These constitute all points at which the problem will be evaluated, such as samples generated through MCS, expected values of optimization variables, most-probable-points in reliability-based design optimization approaches, or collocation grids in generalized polynomial chaos expansion.

% new paragraph
For an example of a Type~II equality constraint, assume that the sum of two length dimensions is required to be a constant value.
If both of these quantities are uncertain, this condition cannot be strictly satisfied. Rather, the constraint may be relaxed or satisfied at its expected value while its standard deviation is minimized.
In this article, we assume that all Type II equality constraints are already relaxed and included in the vector of inequality constraints in Eq.~(\ref{UCCD_ineq}).   

% new paragraph
For the simplicity of notation when deriving the specialized formulations in Sec.~\ref{sec:4}, we define the feasible set of Type~I equality constraints as $\mathcal{E}$:
\begin{align}
    \label{eqn:equalityFeasibility}
    \mathcal{E} = \{ (t,\bm{u}, \bm{\xi}, \bm{p}, \bm{d}) \mid \bm{h}(\cdot) = \bm{0}, \dot{\bm{\xi}}(t) = \bm{f}(\cdot) \}
\end{align}
\noindent
When the inputs to this set are defined probabilistically ($(\tilde{\bm{u}}, \tilde{\bm{\xi}}, \tilde{\bm{d})} \in \mathcal{X}_{\text{stc}}(t)$, and $\tilde{\bm{p}}\in \mathcal{X}_{\text{stc}}$), then $\mathcal{E}$ represent a set in which the analysis-type equality constraints are satisfied almost surely (a.s.) or with the probability of one.

% new paragraph
A fundamental step in formulating the general UCCD problem is identifying the sources of uncertainties that affect ordinary differential equations (ODEs).
When the source of uncertainty is some white-noise, idealized process, such as Wiener and Poisson processes, the resulting differential equations are termed stochastic differential equations (SDEs) \cite{xiu2005high}.
As an example, the motion of electrons in a conductor can be modeled through the Wiener process.
SDEs have been studied extensively and generally require methods based on It\^o and Stratonovich calculus \cite{oksendal2013stochastic}. 
However, for general engineering applications, modeling disturbances as an idealized process is not always sufficient.
Therefore, in this article, we focus on the case where the disturbance vector is a generalized process.
For fuzzy uncertainties, a natural way to model uncertainty propagation in the dynamic system is through fuzzy differential equations (FDEs)~\cite{bayram2018numerical,  higham2001algorithmic, lakshmikantham2004theory}.

%--------------Section 4-----------------%
\xsection{Specialized Formulations }
\label{sec:4}
Based on the previous discussion, it is evident that both uncertainties and problem elements can be represented in different ways\textemdash{}resulting in multiple interpretations of uncertainties with distinct implications on the integrated UCCD solution.
Therefore, it is necessary to formalize some of these interpretations through existing UCCD formulations.

\subsection{Stochastic in Expectation (SE-UCCD)}
\label{subsec:SCCDiE}
Stochastic programming assumes that the probability distributions of the uncertain factors are known.
In these situations, constraints can be modeled in different ways, such as almost surely, in expectation, or in probability~\cite{andrieu2007stochastic}.
Constraints that are described as ``almost surely'' (or ``a.s.'') must be satisfied with the probability of one.
All Type~I equality constraints described in Sec.~\ref{subsec:EC} are a.s.~constraints.
According to Sec.~\ref{par:ExD}, a risk-neutral UCCD problem can be formulated by using the expectation of the objective function and inequality constraints:
\begin{subequations}
\label{Eqn:SCCDE}
\begin{align}
\underset{\tilde{\bm{u}}, \tilde{\bm{\xi}}, \tilde{\bm{p}}}{\textrm{minimize:}}
\quad & o_{\mu} ( t, \tilde{\bm{u}}, \tilde{\bm{\xi}}, \tilde{\bm{p}}, \tilde{\bm{d}})   \label{SCCDE_obj} \\
\textrm{subject to:} \quad
& \bm{g}_{\mu}(t, \tilde{\bm{u}}, \tilde{\bm{\xi}}, \tilde{\bm{p}}, \tilde{\bm{d}}) \leq \bm{0}   \label{SCCDE_ineq}\\
& (t,\tilde{\bm{u}}, \tilde{\bm{\xi}}, \tilde{\bm{p}}, \tilde{\bm{d}}) \in \mathcal{E}
\end{align} 
\end{subequations}
\noindent
Note that in this formulation $(\tilde{\bm{u}}, \tilde{\bm{\xi}}, \tilde{\bm{d}}) \in \mathcal{X}_{\text{stc}}(t)$ and $\tilde{\bm{p}} \in \mathcal{X}_{\text{stc}}$, and $\mathcal{E}$, which was described in Eq.~(\ref{eqn:equalityFeasibility}), represents a set in which analysis-type equality constraints are satisfied almost surely.
Also, the satisfaction of inequality constraints in expectation points to the risk-neutral nature of this formulation.
A lot of real-world CCD problems, however, require explicit risk measures for safety and functionality.
Note that this formulation overlooks some important aspects regarding uncertainty distributions.
For example, the formulation may result in an acceptable mean value but unacceptably low (worst-case) performance.
Reference \cite{Azad2022b} implements a risk-neutral stochastic in expectation UCCD formulation for a simplified strain-actuated solar array system.

\subsection{Stochastic Chance-Constrained (SCC-UCCD)}
\label{subsec:SccCCD}
\begin{figure}[t]
\centering
\begin{subfigure}[t]{0.495\columnwidth}
\centering
\includegraphics[scale=0.38]{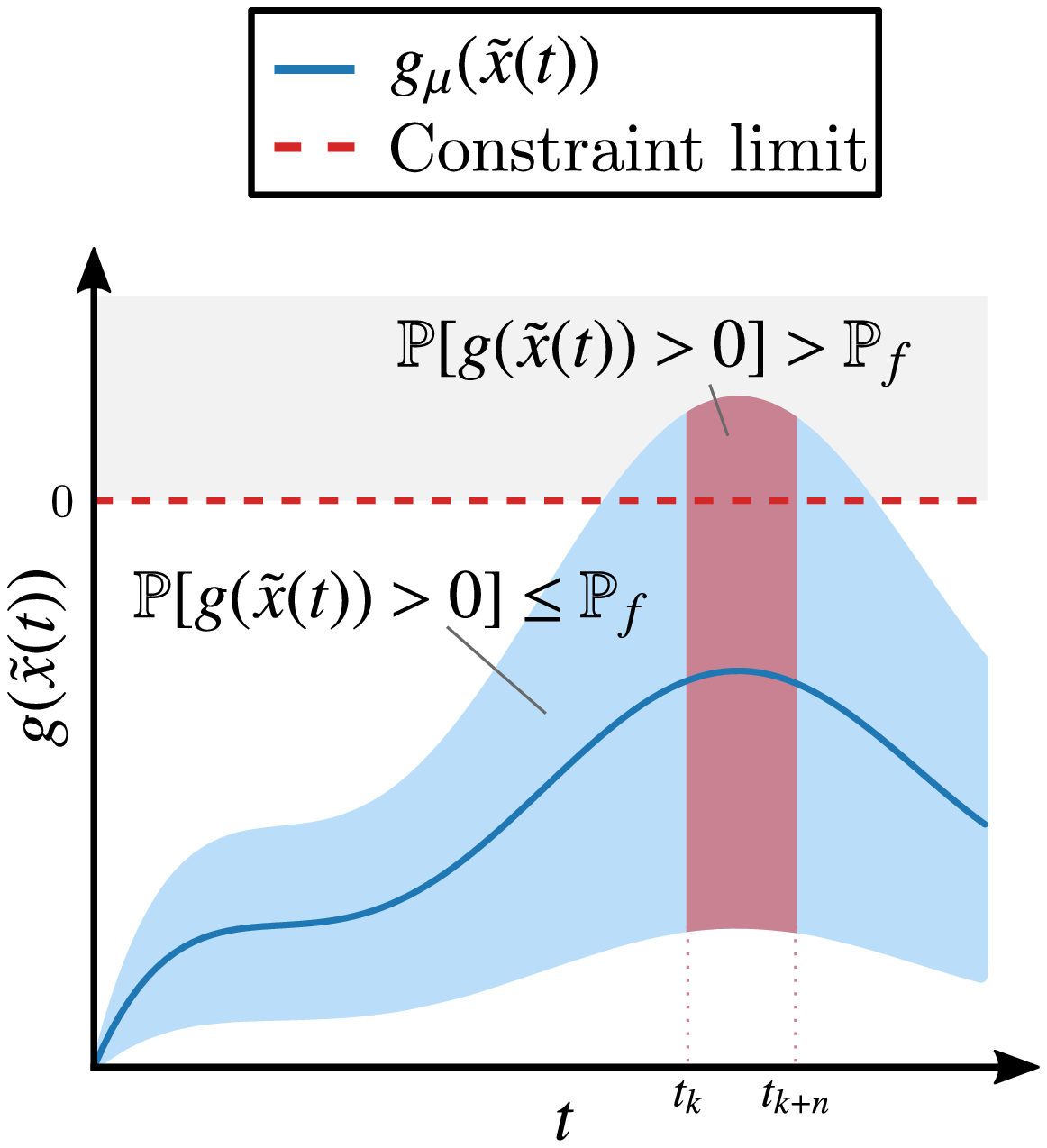}
\caption{Stochastic Chance-Constraint.}
\label{subfig:4a}
\end{subfigure}%
\hspace{0.01\columnwidth}%
\begin{subfigure}[t]{0.495\columnwidth}
\centering
\includegraphics[scale=0.38]{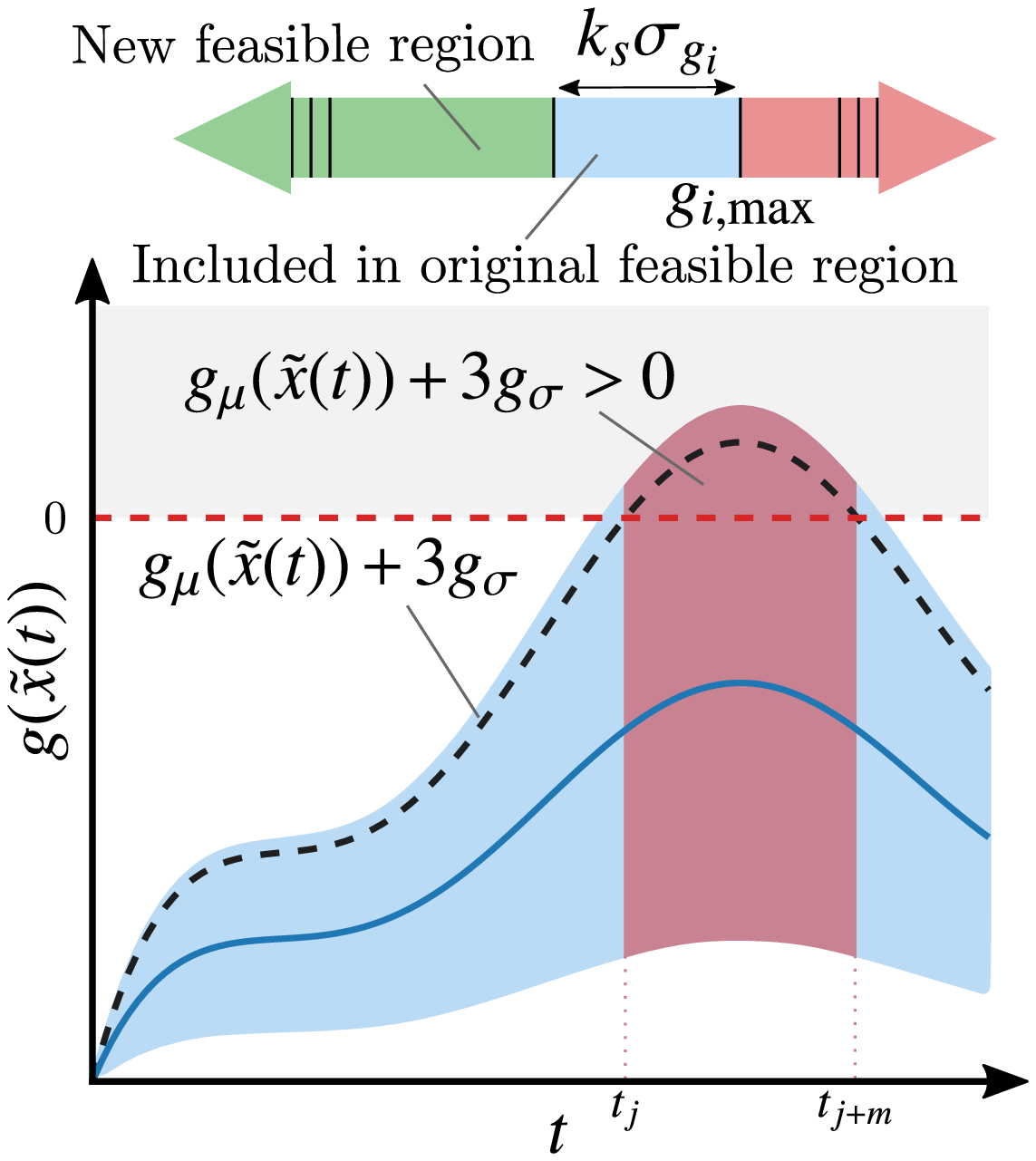}
\caption{Probabilistic Robust.}
\label{subfig:4b}
\end{subfigure}%
\hspace{0.015\textwidth}
\caption{Illustration of uncertain probabilistic constraints (a)~stochastic path constraint with prescribed failure probability of $\mathbb{P}_{f}$ and, (b)~ probabilistic robust constraint interpretation with constraint shift index $k_{s}=3$.}
\label{fig:cons}
\end{figure}

Problems with probabilistic inequality constraints are generally referred to as chance-constrained programming. They are ubiquitous in various research fields, such as reliability-based design optimization (RBDO) and trajectory optimization.
Recently, novel UCCD formulations based on RBDO have been developed in Refs.~\cite{azad2020single,  cui2020comparative}.
Here we introduce a more general chance-constrained formulation referred to as stochastic chance-constrained UCCD. The problem formulation is described as:  
\begin{subequations}
 \label{Eqn:CCCCD}
 \begin{align}
 \underset{\tilde{\bm{u}}, \tilde{\bm{\xi}}, \tilde{\bm{p}}}{\textrm{minimize:}}
 \quad & o_{\mu} ( t, \tilde{\bm{u}}, \tilde{\bm{\xi}}, \tilde{\bm{p}}, \tilde{\bm{d}})  \label{CCCCD_obj} \\
 \textrm{subject to:} \quad
& \mathbb{P}[g_{i}(t, \tilde{\bm{u}}, \tilde{\bm{\xi}}, \tilde{\bm{p}}, \tilde{\bm{d}}) > 0] \leq \mathbb{P}_{f,i} \ \ \ i = 1, \dots, n_g \label{CCCCD_ineq} \\
& (t,\tilde{\bm{u}}, \tilde{\bm{\xi}}, \tilde{\bm{p}}, \tilde{\bm{d}}) \in \mathcal{E}
\end{align} 
\end{subequations}
\noindent
Again, in this formulation we have $(\tilde{\bm{u}}, \tilde{\bm{\xi}}, \tilde{\bm{d}}) \in \mathcal{X}_{\text{stc}}(t)$ and $\tilde{\bm{p}} \in \mathcal{X}_{\text{stc}}$.
Analysis-type equality constraints are satisfied almost surely, and the probabilistic representation of inequality constraints ensures that they are satisfied with a given target reliability of $1-\mathbb{P}_{f,i}$.
The stochastic interpretation of path constraints is further illustrated in Fig.~\ref{subfig:4a}.
In this figure, blue areas have failure probabilities that do not exceed $\mathbb{P}_{f}$, while red regions violate the constraint with probabilities higher than $\mathbb{P}_{f}$.
When used only with open-loop control, the above formulation may lead to conservative trajectories.
This is because, in practice, feedback controllers are often implemented for such systems and have the capacity to compensate for some of these uncertainties.
However, when only open-loop control is considered, Eq.~(\ref{Eqn:CCCCD}) often neglects the possible role of feedback controller at the time of implementation~\cite{lew2020chance}.
Therefore, closing the control loop in such UCCD problems may entail improvements in performance and cost.
A chance-constrained stochastic, nonlinear control strategy for motion planning of robotic systems is introduced in Ref.~\cite{nakka2022trajectory}.
Furthermore, a stochastic chance-constrained implementation for UCCD case studies, using concepts from reliability-based design optimization, are presented in Refs.~\cite{azad2020single, cui2020comparative}.

% Figure 4

\subsection{Probabilistic Robust (PR-UCCD)}
\label{subsubsec:pra}
If we assume that the decision-maker has some knowledge about the probabilistic behavior of uncertainties, a robust interpretation, which is often credited to Genichi Taguchi~\cite{taguchi1986introduction}, may be utilized.
In this interpretation, robustness is defined as the reduced sensitivity of the objective function and constraints to variations in uncertain quantities.
Robustness measures commonly used with this interpretation are the expectancy and dispersion, which were described in Secs.~\ref{par:ExD} and \ref{par:HmD}, respectively, and are commonly used together in a multiobjective optimization problem to find a compromise solution.
Thus, methods from robust multiobjective optimization are generally used with such formulations \cite{ide2016robustness}.
The PR-UCCD problem can be written as:  
\begin{subequations}
 \label{Eqn:RCCD_p1}
 \begin{align}
 \underset{\tilde{\bm{u}}, \tilde{\bm{\xi}}, \tilde{\bm{p}}}{\textrm{minimize:}} 
 \ \ \ & \alpha_w o_{\mu}(t,\tilde{\bm{u}}, \tilde{\bm{\xi}},\tilde{\bm{p}}, \tilde{\bm{d}}) + (1-\alpha_w) o_{\sigma}(t,\tilde{\bm{u}}, \tilde{\bm{\xi}},\tilde{\bm{p}}, \tilde{\bm{d}})   \label{RCCD_obj_p1} \\
 \textrm{subject to:} \ \ \
& \bm{g}_{\mu}(t, \tilde{\bm{u}}, \tilde{\bm{\xi}}, \tilde{\bm{p}}, \tilde{\bm{d}}) + k_{s} \bm{g}_{\sigma}(t,\tilde{\bm{u}}, \tilde{\bm{\xi}}, \tilde{\bm{p}}, \tilde{\bm{d}}) \leq \bm{0} \label{RCCD_ineq_p1} \\
& (t,\tilde{\bm{u}}, \tilde{\bm{\xi}}, \tilde{\bm{p}}, \tilde{\bm{d}}) \in \mathcal{E}
\end{align}
\end{subequations}
\noindent
where $(\tilde{\bm{u}}, \tilde{\bm{\xi}}, \tilde{\bm{d}}) \in \mathcal{X}_{\text{stc}}(t)$ and $\tilde{\bm{p}} \in \mathcal{X}_{\text{stc}}$.
In addition, $\alpha_w$ and $(1-\alpha_w)$ are weights associated with the multiobjective optimization problem.
In the above formulation, a constraint shift index $k_{s}$, selected by the designer, is used to reduce the feasibility region of constraints.
This approach practically moves the optimal solution away from constraint boundaries but does not always offer a probabilistic interpretation. 
Alternatively, the problem can be formulated as:
\begin{subequations}
 \label{Eqn:RCCD_p2}
 \begin{align}
 \underset{\tilde{\bm{u}}, \tilde{\bm{\xi}}, \tilde{\bm{p}}}{\textrm{minimize:}} 
 \quad & \alpha_w o_{\mu}(t,\tilde{\bm{u}}, \tilde{\bm{\xi}},\tilde{\bm{p}}, \tilde{\bm{d}})  + (1-\alpha_w)o_{\sigma}(t,\tilde{\bm{u}}, \tilde{\bm{\xi}},\tilde{\bm{p}}, \tilde{\bm{d}} )  \label{RCCD_obj_p2} \\
 \textrm{subject to:} \quad
& \bm{g}_{\mu}(t, \tilde{\bm{u}}, \tilde{\bm{\xi}}, \tilde{\bm{p}}, \tilde{\bm{d}})  \leq \bm{0} \label{RCCD_ineq1_p2} \\
& \bm{g}_{\sigma}(t, \tilde{\bm{u}}, \tilde{\bm{\xi}}, \tilde{\bm{p}}, \tilde{\bm{d}}) - \bm{\sigma}_{a} \leq \bm{0} \label{RCCD_ineq2_p2}\\
& (t,\tilde{\bm{u}}, \tilde{\bm{\xi}}, \tilde{\bm{p}}, \tilde{\bm{d}}) \in \mathcal{E}
\end{align}
\end{subequations}  
\noindent
where $\sigma_{a}$ is the allowable standard deviation for $\bm{g}(\cdot)$. 
In this formulation, the uncertain inequality constraints are satisfied at their expected value, and their corresponding standard deviation (or variance) is below the allowable limit.
Probabilistic robust path constraints are further illustrated in Fig.~\ref{subfig:4b}, under the assumption of {a} Gaussian distribution with zero skew.
In the top part of the illustration, the reduced feasible space for constraints with simple bounds is demonstrated, while the bottom shows the $3g_{\sigma}$ bound for an arbitrary path constraint.
One of the limitations of the PR-UCCD formulation is that all of the scenarios that differ from the expectation are penalized, regardless of performance.
In other words, the formulation penalizes the superior (i.e.,~better than the mean value) and the poor performance (i.e.,~worse than the mean value) simultaneously.
A more detailed discussion on the implications of using dispersion as a robustness measure is provided in Sec.~\ref{subsec:PR_robust}.
References \cite{azad2021robust, xun2022chance} use the probabilistic UCCD formulation for the UCCD problem of a hybrid electric vehicle powertrain and a fuel cell hybrid electric truck, respectively. 

% new paragraph
A major challenge associated with the probabilistic formulations presented so far is that obtaining distributional information about the uncertain factors is not always viable.
In addition, even if this information can be estimated, the resulting formulation is generally computationally intractable~\cite{shapiro2005complexity}.
The first challenge is generally addressed by using concepts from robust optimization, which is discussed next.

\subsection{Worst-Case Robust (WCR-UCCD)}
\label{subsec:wcRCCD}
Robustness in UCCD is motivated by the fact that when a solution to a deterministic CCD problem exhibits large sensitivities to perturbations in problem parameters, it becomes highly infeasible and impractical.
This issue has been traditionally addressed by robust control, as well as robust design optimization communities in disparate efforts.
However, to utilize the full synergistic performance potential of UCCD, both plant design and control system domains must be explored simultaneously in a balanced way.
While robust UCCD has only been investigated in a handful of studies \cite{azad2020robust, azad2021robust, nash2021robust}, there's a need for  practical formulations and interpretations of robustness in UCCD problems.  
In this section, we first describe robustness and its associated worst-case realization and then introduce the WCR-UCCD formulation.

% Figure 5
\begin{figure}[t]
\centering
\includegraphics[scale=0.65]{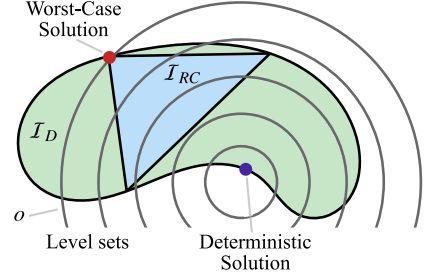}
\caption{Illustration of the worst-case solution in context of the constraint feasible space and uncertainty sets.}
\label{fig:wcr}
\end{figure}

\subsubsection{Robust Interpretation.\label{par:WcRRI}}~In its most common interpretation, a solution is robust if it remains feasible for all of the realizations of uncertainty within the uncertainty set.
This notion naturally leads to the definition of an equivalent deterministic formulation that is referred to as the robust counterpart (RC).
Utilizing the epigraph representation of the objective function introduced in Sec.~\ref{subsec:OF}, the RC of the general UCCD problem can then be formulated as:
\begin{equation}  
\label{Eqn:RCCD1}
 \begin{aligned}
 \underset{{\hat{\bm{u}}}, {\hat{\bm{\xi}}}, {\hat{\bm{p}}}, v}{\textrm{minimize:}} \quad & v  \\
 \textrm{subject to:} \quad & 
   \begin{rcases}
 \bm{g} ( t, {\bm{u}}, {\bm{\xi}}, {\bm{p}}, {\bm{d}})  \leq {\bm{0}}   \\ 
  o(t, {\bm{u}}, {\bm{\xi}}, {\bm{p}}, {\bm{d}}) - v \leq 0  \\
  {(t,{\bm{u}}, {\bm{\xi}}, {\bm{p}}, {\bm{d}}) \in \mathcal{E}}
\end{rcases} %\notag
 \begin{split}
& {\forall~({\bm{u}}, {{\bm{\xi}},} {\bm{d}} ) \in \mathcal{R}_{t}(\hat{\bm{q}}_{t})}\\
& {\forall~{\bm{p}} \in \mathcal{R}(\hat{\bm{q}})}
\end{split} 
\end{aligned} 
\end{equation}
\noindent
where $(\hat{\bm{u}}, \hat{\bm{\xi}}, \hat{\bm{d}}) \in \mathcal{R}_{t}(\hat{q}_{t})$ and $\hat{\bm{p}} \in \mathcal{R}(\hat{q})$ are nominal set parameters that result in the smallest value of the objective function $v$ that can simultaneously satisfy all of the constraints for all uncertainty realizations within the set.  
Note that the entirety of uncertainty sets are imposed on the constraint feasible space.
Depending on the properties of our uncertainty sets, this may result in a finite or infinite number of hard constraints.
In the remainder of this section, we assume that the new deterministic optimization variable $v$ is included in the vector of time-independent optimization variables ${\bm{p}}$, and the new inequality constraint is included in $\bm{g}(\cdot)$. 

If Eq.~(\ref{Eqn:RCCD1}) is to be satisfied for every realization of uncertainties, then
$\mathcal{R}_{t}(\hat{\bm{q}_{t}}) \times \mathcal{R}(\hat{\bm{q}})$ must be contained within the constraint feasibility set.
Mathematically, the constraint feasible space can be defined as: 
\begin{gather}
\mathcal{I}_{D} \coloneqq  \Bigl\{ ({\bm{u}}, \bm{\xi}, {\bm{p}}, {\bm{d}}) \mid \bigl\{ \left \{\bm{g}(\cdot)\leq \bm{0} \right \} \cap \mathcal{E}(\cdot) \bigl\} \Bigl\} 
\end{gather}
\noindent
and the feasible space of the RC problem can be described as:
\begin{gather}
\mathcal{I}_{RC} \coloneqq \Bigl\{
({\bm{u}}, \bm{\xi}, {\bm{p}}, {\bm{d}}) \in  \bigl\{ \{\mathcal{R}_{t}(\hat{\bm{q}}_{t}) \times \mathcal{R}(\hat{\bm{q}}) \} \cap \mathcal{I}_{D} \bigl\} \Bigl\} 
\end{gather}
\noindent
where $\mathcal{I}_{RC} \subseteq \mathcal{I}_{D}$.
This definition, which is required for the solution of Eq.~(\ref{Eqn:RCCD1}), sheds some light on some of the considerations in constructing uncertainty sets for practical robust implementations. 
For some bounded uncertainty set, this notion is conceptually illustrated for $\mathcal{I}_{RC}$ and $\mathcal{I}_{D}$ in Fig.~\ref{fig:wcr}.

\subsubsection{Worst-Case Robust Interpretation.\label{par:WcRWcI}}~When the uncertainty set is infinite, Eq.~(\ref{Eqn:RCCD1}) is a semi-infinite problem where there's a finite number of decision variables and an infinite number of constraints.
Generally, this RC problem is large, intractable, and difficult to solve.
For instance, the RC of a linear optimization problem is typically a nonlinear optimization problem.
Despite such difficulties, the robust interpretation offers a certain relative simplicity and computational viability compared to other interpretations, making it a valuable tool for understanding and addressing uncertainties in many engineering problems, including UCCD.

% new paragraph
One approach to deal with this semi-infinite problem is to replace the infinite uncertainty set with a finite subset or a sequence of successively refined grids~\cite{hettich1993semi}.
A more constructive approach, however, is to replace semi-infinite constraints with the solution of the constraint maximization problem.
To understand this idea, we draw an analogy from the game theory literature.
Assume that the optimizer has a natural adversarial opponent \cite{bryson2018applied, diehl2008numerical}.
Therefore, for every decision the optimizer makes, the adversarial opponent makes a decision (over uncertainties) to disturb constraints as strongly as possible.
This notion leads to the realization of worst-case uncertainties and, consequently, the concept of $\min-\max$, or minimax robust formulation, which was briefly introduced in Eq.~(\ref{Eqn:wcobjIneq}).

\subsubsection{WCR-UCCD Formulation.\label{par:WcRF}}~To adopt the WCR interpretation for UCCD, we need to differentiate between the decision space of the optimizer and the decision space of the adverse player.
In addition to the analysis-type feasibility space, which affects both players, the adverse player is restricted in its decisions to uncertainties contained within $\mathcal{R}_{t}(\hat{\bm{q}}_{t}) \times \mathcal{R}(\hat{\bm{q}})$.
The WCR-UCCD problem is now formulated such that the deterministic objective function $v$ is minimized over the set of optimizations variables $[\hat{\bm{u}}, \hat{\bm{\xi}}, \hat{\bm{p}}]$, subject to constraint maximization problems, Type~I feasibility set, and (potentially) additional feasibility constraints:
\begin{subequations}
 \label{Eqn:RCCDc}
 \begin{align}
\underset{\hat{\bm{u}}, {\hat{\bm{\xi}}},\hat{\bm{p}}}{\textrm{minimize:}}  \quad & v \\
\textrm{subject to:} \quad
&  \Phi_{i}(t, \hat{\bm{u}},\hat{\bm{\xi}},\hat{\bm{p}},\hat{\bm{d}}) \leq 0 ~~ \textrm{for}~~i = 1, \dots , n_g \label{RCCDc_ineq}\\
& (t,\hat{\bm{u}}, \hat{\bm{\xi}}, \hat{\bm{p}}, \hat{\bm{d}}) \in \mathcal{E} \label{RCCDc_eq}\\
& \bm{\psi}(\hat{\bm{u}}, \hat{\bm{\xi}}, \hat{\bm{p}}, \hat{\bm{d}}) \leq \bm{0}
\end{align}
\end{subequations}
\noindent
where $(\hat{\bm{u}},\hat{\bm{\xi}}, \hat{\bm{d}}) \in \mathcal{R}_{t}(\hat{\bm{q}}_{t}) \subseteq \mathcal{X}_{\text{crisp}}(t)$ and $\hat{\bm{p}} \in \mathcal{R}(\hat{\bm{q}})$ are inputs to the inner-loop optimization problem for all $n_g$ inequality constraints. 
Equation~(\ref{RCCDc_eq}) ensures that the nominal set parameters satisfy the analysis-type equality constraints.
$\bm{\psi}(\cdot)$ are optional additional feasibility constraints, similar to the ones used in Ref.~\cite{herber2019nested}. 
The inner-loop maximization problem $\Phi_{i}(\cdot)$ is: 
\begin{subequations}
 \label{Eqn:RCCDd}
 \begin{align}
\underset{{\bm{u}}, {\bm{\xi}}, {\bm{p}}, {\bm{d}}}{\textrm{maximize:}}  \quad  & g_{i}(t, {\bm{u}}, {\bm{\xi}}, {\bm{p}}, {\bm{d}})  \label{RCCDd_obj} \\
\textrm{subject to:} \quad 
& (t,{\bm{u}}, {\bm{\xi}}, {\bm{p}}, {\bm{d}}) \in \mathcal{E} \label{RCCDd_eq}\\
& (\bm{u}, \bm{\xi}, \bm{d}) \in \mathcal{R}_{t}(\hat{\bm{q}}_{t}), ~
\bm{p} \in \mathcal{R}(\hat{\bm{q}}) 
\end{align}
\end{subequations}
\noindent
where $(\bm{u}, \bm{\xi}, \bm{\bm{d}}) \in \mathcal{R}_{t}(\hat{\bm{q}}_{t}) \subseteq \mathcal{X}_{\text{crisp}}(t)$ and $\bm{p} \in \mathcal{R}(\hat{\bm{q}}) \subseteq \mathcal{X}_{\text{crisp}}$ are the worst-case combination of uncertainties belonging to their associated sets for constraint $i$.
This inner-loop optimization problem attempts to maximize $g_{i}$ by selecting the worst-case combination of uncertainties, subject to all of the Type I equality constraints and the definition of the uncertainty sets.
The feasibility sets associated with the inner-loop and outer-loop problem structure require special considerations similar to the ones described in Ref.~\cite{herber2019nested}.   
This WCR-UCCD formulation, which presents the broad case of independent uncertainties within all problem elements, is decomposed such that the optimization problem of the decision-maker is formulated in the outer loop, and the optimization of the adversarial player is formulated in the inner loop.
Depending on the problem at hand, other coordination strategies may also be used.
This interpretation of robustness has been used along with a model predictive control strategy to find a robust UCCD solution of an aircraft thermal management system in Ref.~\cite{nash2021robust}.
In addition, Ref.~\cite{Azad2022b} compares the worst-case robust UCCD solution of a simplified strain-actuated solar array to that of the stochastic in expectation UCCD.

\subsection{Fuzzy Expected Value (FE-UCCD)}
\label{subsubsec:pose}
When uncertainties in UCCD are represented as fuzzy variables and processes, the UCCD problem can be formulated using a fuzzy expected-value model.
The challenge is to choose optimization variables such that the objective function, which is related to some fuzzy processes (through fuzzy differential equations), is optimized.  
Here, we use the expected-value model~\cite{zhu2009fuzzy, liu2002toward}:
\begin{subequations}
 \label{Eqn:PCCCCDa}
 \begin{align}
 \underset{\tilde{\bm{u}}, \tilde{\bm{\xi}}, \tilde{\bm{p}}}{\textrm{minimize:}}   
 \quad & \mathbb{E}[o(t,\tilde{\bm{u}}, \tilde{\bm{\xi}},\tilde{\bm{p}}, \tilde{\bm{d}})] \label{PCCCCDa_obj} \\
 \textrm{subject to:} \quad 
& \mathbb{E}[\bm{g}(t, \tilde{\bm{u}}, \tilde{\bm{\xi}}, \tilde{\bm{p}}, \tilde{\bm{d}}, \tilde{\bm{w}})] \leq \bm{0} \label{PCCCCDa_ineq} \\
 & (t,\tilde{\bm{u}}, \tilde{\bm{\xi}}, \tilde{\bm{p}}, \tilde{\bm{d}}) \in \mathcal{E}
\end{align}
\end{subequations} 
\noindent
where we note that in this formulation $(\tilde{\bm{u}}, \tilde{\bm{\xi}}, \tilde{\bm{d}}) \in \mathcal{X}_{\text{fuzzy}}(t)$ and $\tilde{\bm{p}} \in \mathcal{X}_{\text{fuzzy}}$.
In this formulation, $\mathcal{E}$ refers to the feasibility set of analysis-type equality constraints that now contain fuzzy differential equations.
Reference \cite{soni2012optimal} uses the fuzzy expected value model for optimal pricing and inventory policies.

\subsection{Possibilistic Chance-Constrained (PCC-UCCD)}
\label{subsubsec:posc}
As opposed to FE-UCCD, which is a risk-neutral formulation, PCC-UCCD utilizes a possibility measure to hedge against uncertainties.
This measure ensures that fuzzy constraints hold within a given confidence threshold \cite{liu2002toward}. 
The possibility-based chance-constrained UCCD formulation can be written as:
\begin{subequations}
 \label{Eqn:PCCCCD}
 \begin{align}
 \underset{\tilde{\bm{u}}, \tilde{\bm{\xi}}, \tilde{\bm{p}}}{\textrm{minimize:}}   
 \quad & \mathbb{E}[o(t,\tilde{\bm{u}}, \tilde{\bm{\xi}},\tilde{\bm{p}}, \tilde{\bm{d}})] \label{PCCCCD_obj} \\
 \textrm{subject to:} \quad
& \mathbb{P}\mathbb{O}\mathbb{S}[g_{i}(t, \tilde{\bm{u}}, \tilde{\bm{\xi}}, \tilde{\bm{p}}, \tilde{\bm{d}}) > 0] \leq \mathbb{P}\mathbb{O}\mathbb{S}_{f,i} \label{PCCCCD_ineq} \\
 & (t,\tilde{\bm{u}}, \tilde{\bm{\xi}}, \tilde{\bm{p}}, \tilde{\bm{d}}) \in \mathcal{E}
\end{align}
\end{subequations} 

\noindent
where $\mathbb{P}\mathbb{O}\mathbb{S}_{f,i}$ is the target possibility of failure of constraint $i$, and $(\tilde{\bm{u}}, \tilde{\bm{\xi}}, \tilde{\bm{d}}) \in \mathcal{X}_{\text{fuzzy}}(t)$ and $\tilde{\bm{p}} \in \mathcal{X}_{\text{fuzzy}}$ and $\dot{\tilde{\bm{\xi}}} - \bm{f}(\cdot) = 0$ contained in $\mathcal{E}$ are now fuzzy differential equations.
A possibilistic chance-constrained formulation for {a} unit commitment problem involving demand response, electric vehicles, and wind power is presented in Ref.~\cite{zhang2015fuzzy}.

%--------------Section 5-----------------%
\xsection{Discussion}
\label{sec:5}
With various formulations now defined, we discuss several aspects of them in more detail, focusing on their connections and existing research.

\subsection{Norm-Induced Uncertainty Sets}
\label{subsec:CUS}
The worst-case robust formulation, introduced in Sec.~\ref{par:WcRF}, is directly related to the choice of $\mathcal{R}_{t}(\hat{\bm{q}}_{t})$ and $\mathcal{R}(\hat{\bm{q}})$.
In robust optimization, these uncertainty sets are generally defined according to some norm.
Using only the notation for time-independent variables, these norm-induced uncertainty sets are mathematically defined as:
\begin{align}
    \label{Eqn:rs}
    \mathcal{N} \coloneqq  & \left \{ \bm{q}~\mid \bm{z} \left( \hat{\bm{q}} - \bm{q} \right )  \leq \bm{\eta}_{q} \right \} 
\end{align}
\noindent 
where $\bm{z}(\cdot)$ is a specified function chosen to represent the geometry of the uncertainty set, often through an applied norm such as $\ell_{1}$, $\ell_{2}$, $\ell_{p}$, $D$, CVaR, etc. \cite{rahal2021norm}.
The resulting uncertainty sets may have different shapes and geometries, such as box, ellipsoidal, polyhedron, etc.
The size of the uncertainty sets, which is also a modeling choice, is prescribed through $\bm{\eta}_{q}$, which is included in the vector of problem data.
Through these parameters, the decision-maker has the advantage of leveraging the size and structure of the uncertainty set to benefit from different properties of the resulting sets~\cite{gorissen2015practical, bertsimas2011theory, ben2009robust, ben1998robust}.
As an example, a simple box uncertainty set for plant optimization variables can be defined as $\bm{z}(\hat{\bm{p}}) = \vert\hat{\bm{p}} - {\bm{p}}\vert$
and  $\bm{\eta}_{p} = \bm{\Delta} p$.

% new paragraph
Note that if the size of the selected set compared to the reality of the uncertain phenomenon is too large or too small, it might result in a solution that is too conservative or high-risk, respectively.
To address this issue, one may attempt to optimally leverage the uncertainty set's size, shape, and structure to obtain a meaningful solution for a given metric.
This requires that the uncertainty sets are treated as additional optimization variables, leading to the concept of adjustable uncertainty sets as described in Refs.~\cite{zhang2017robust, kim2018robust}.
Robust unit commitment with adjustable uncertainty sets for uncertain wind generation is discussed in Ref.~\cite{wang2016robust}.

\subsection{Linking Stochastic and Worst-Case Robust Formulations}
\label{subsec:Connection}
Different forms of uncertainty representation lead to different interpretations and, therefore, problem formulations.
Specifically, in SCC-UCCD, it is assumed that the probability distribution of uncertainties is known or can be estimated.
In contrast, the WCR-UCCD assumes that uncertainties belong to a crisp set and no probabilistic information is available.
Therefore, while the SCC-UCCD gives a probabilistic measure to quantify the risks associated with constraint violation, the robust UCCD cannot offer such a measure.
Nevertheless, strict satisfaction of (infinitely many) hard constraints in WCR-UCCD in Eq.~(\ref{Eqn:RCCD1}) (when an appropriately sized/shaped uncertainty set is selected) is equivalent (in the limit) to the satisfaction of probabilistic constraints in SCC-UCCD with an infinitesimally small failure probability.

% new paragraph
In addition, in modern robust approaches, the size and geometry of the uncertainty sets may be leveraged to adjust the associated risk.
For instance, increasing the size of the uncertainty set in the WCR-UCCD increases the number of constraints that need to be satisfied in Eq.~(\ref{Eqn:RCCD1}), which is equivalent to reducing the probability of failure $\mathbb{P}_{f}$ in SCC-UCCD formulations.
Finally, Refs.~\cite{beyer2007robust, bertsimas2011theory} offer probabilistic interpretations of robust formulations, which practically bridge the gap between the minimax interpretation of robust formulations and the probabilistic interpretation of stochastic chance-constrained problems.
This interpretation leads to the notion of probabilistic guarantees for robust optimization problems and seeks to connect robust feasibility to the probability of feasibility.
Consequently, even when the underlying distribution is known, benefits from the tractability of robust formulations may compel one to use such probabilistic guarantees in robust formulations instead of using stochastic ones.
Such probabilistic guarantees may be computed a priori as a function of the structure and size of the uncertainty set and lead to the notion of a budget of uncertainty~\cite{bertsimas2011theory}.

\subsection{Robustness in the PR-UCCD Formulation}
\label{subsec:PR_robust}

% Figure 6
\begin{figure*}[t]
\centering
\begin{subfigure}[t]{0.25\textwidth}
\centering
\includegraphics[scale=0.3]{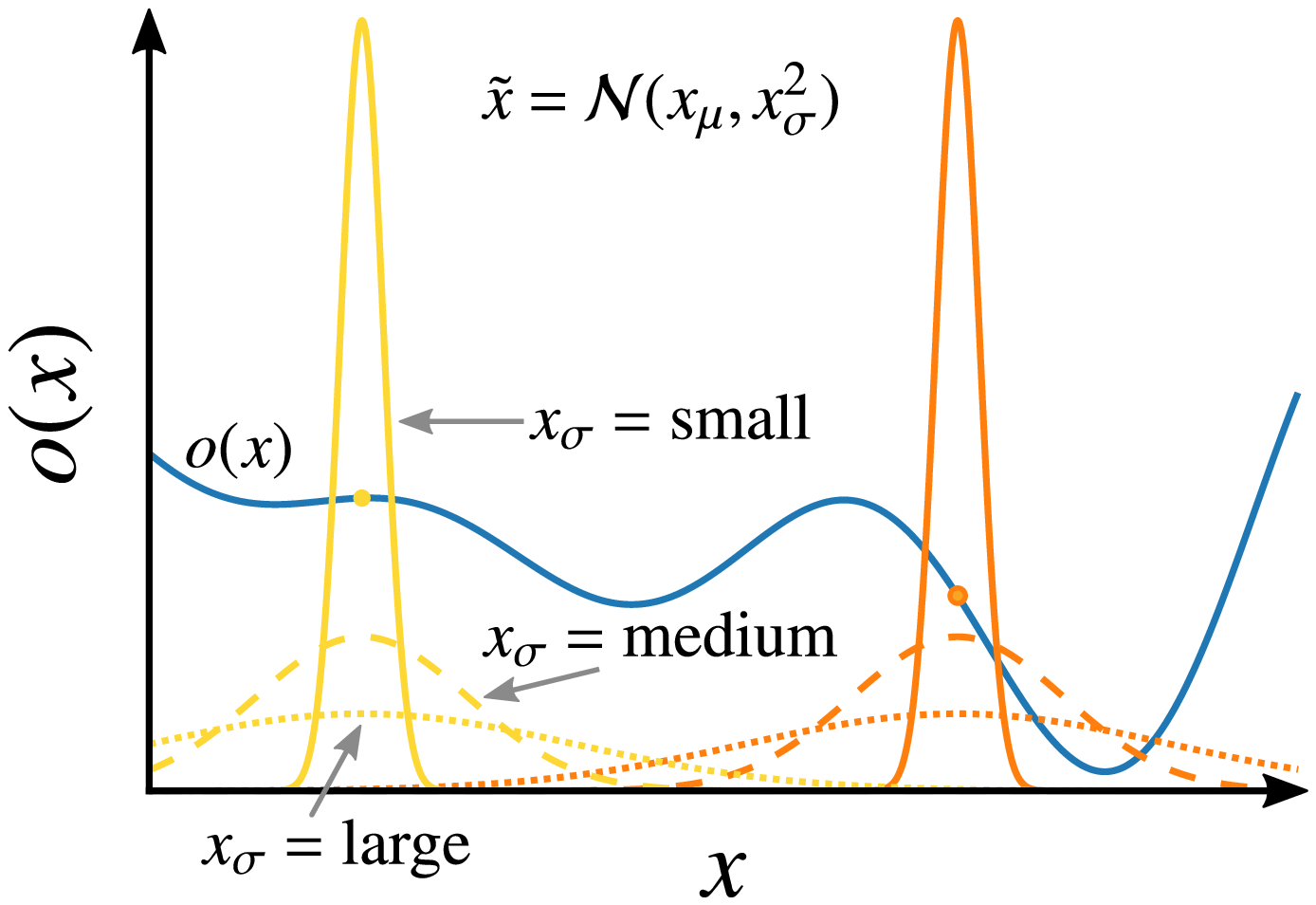}
\caption{Deterministic $o(x)$.}
\label{subfig:deto}
\end{subfigure}%
\begin{subfigure}[t]{0.25\textwidth}
\centering
\includegraphics[scale=0.3]{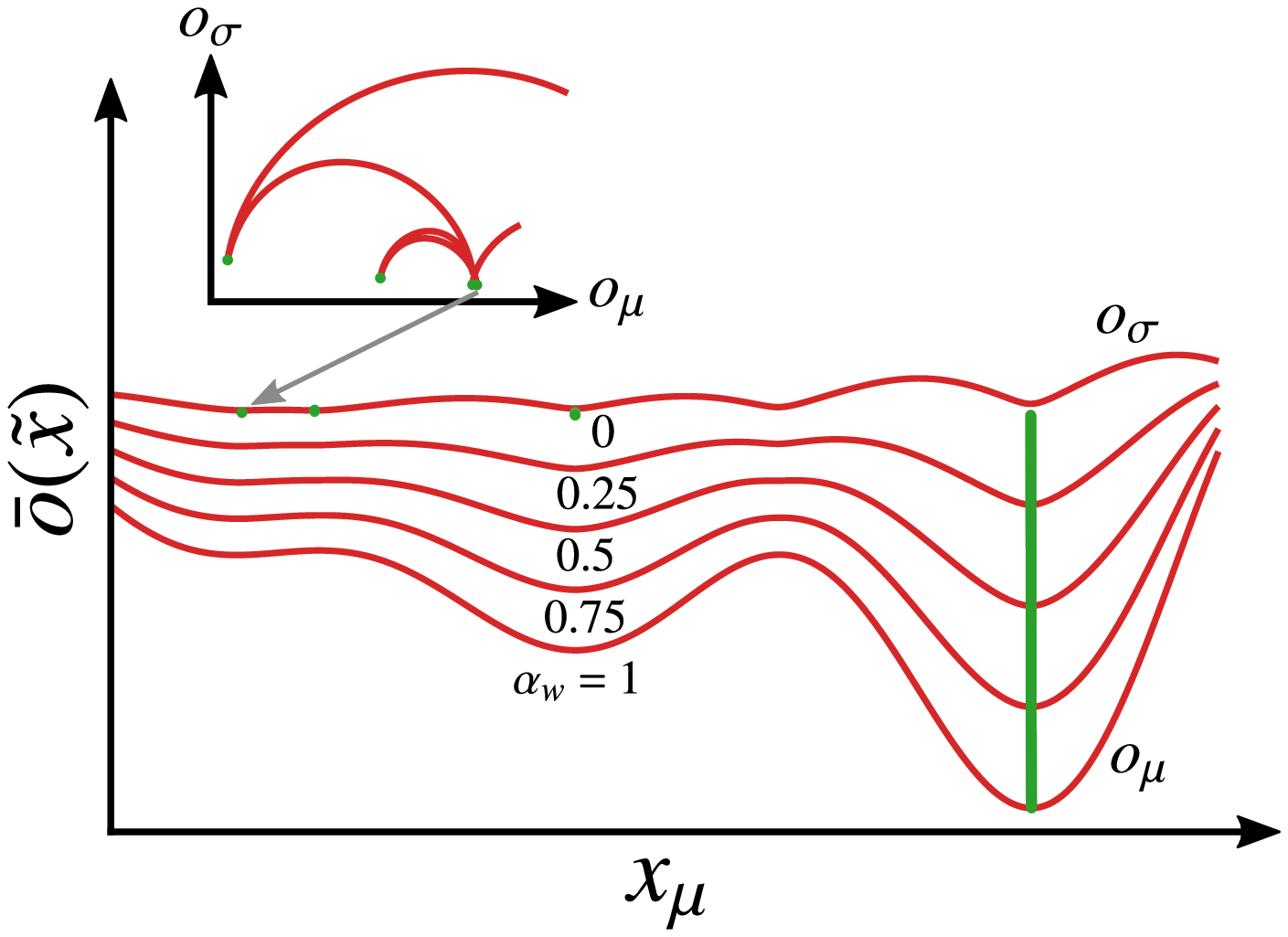}
\caption{Small uncertainty.}
\label{subfig:PRo_S}
\end{subfigure}%
\begin{subfigure}[t]{0.25\textwidth}
\centering
\includegraphics[scale=0.3]{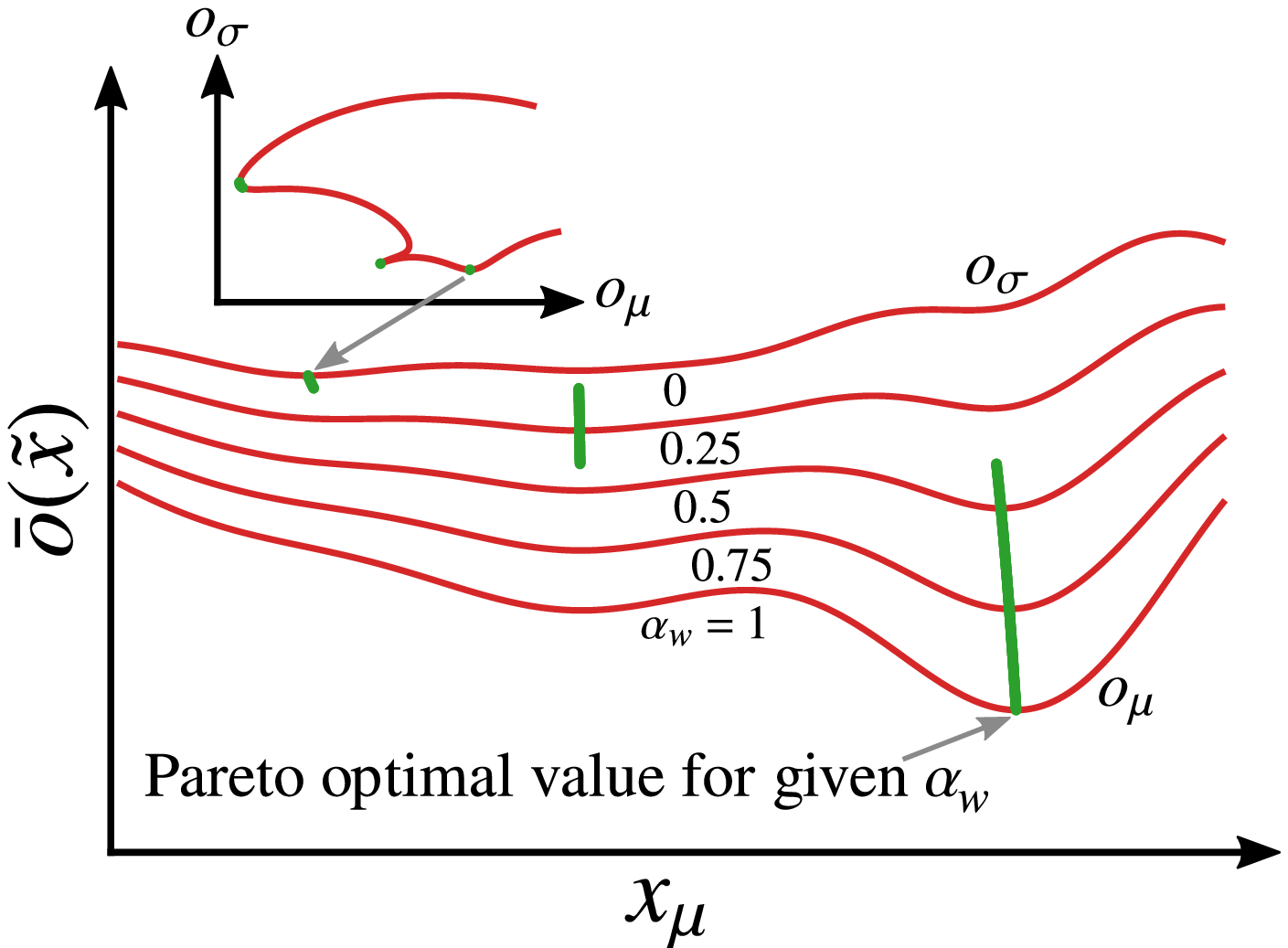}
\caption{Medium uncertainty.}
\label{subfig:PRo_M}
\end{subfigure}%
\begin{subfigure}[t]{0.25\textwidth}
\centering
\includegraphics[scale=0.3]{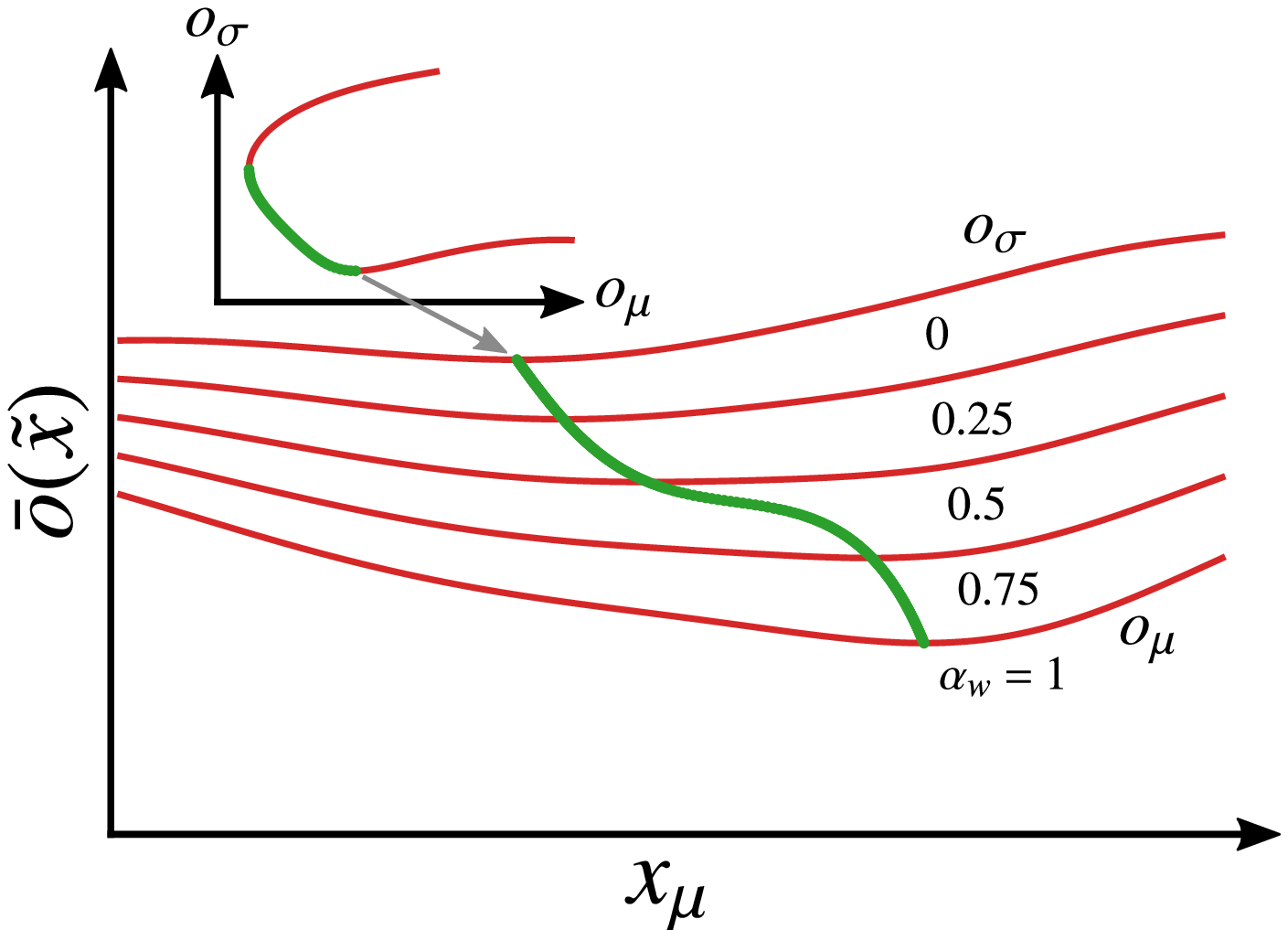}
\caption{Large uncertainty.}
\label{subfig:PRo_L}
\end{subfigure}
\captionsetup[figure]{justification=centering}
\caption{Illustration of a Pareto trade-offs in the probabilistic robust objective function for several different levels of uncertainty ($x_\sigma$) of a single uncertain variable $\tilde{x} = \mathcal{N}(x_{\mu},x^2_{\sigma})$.}
\label{fig:PRObjective}
\end{figure*}

% Figure 7
\begin{figure*}[t]
\centering
\begin{subfigure}[t]{0.25\textwidth}
\centering
\includegraphics[scale=0.3]{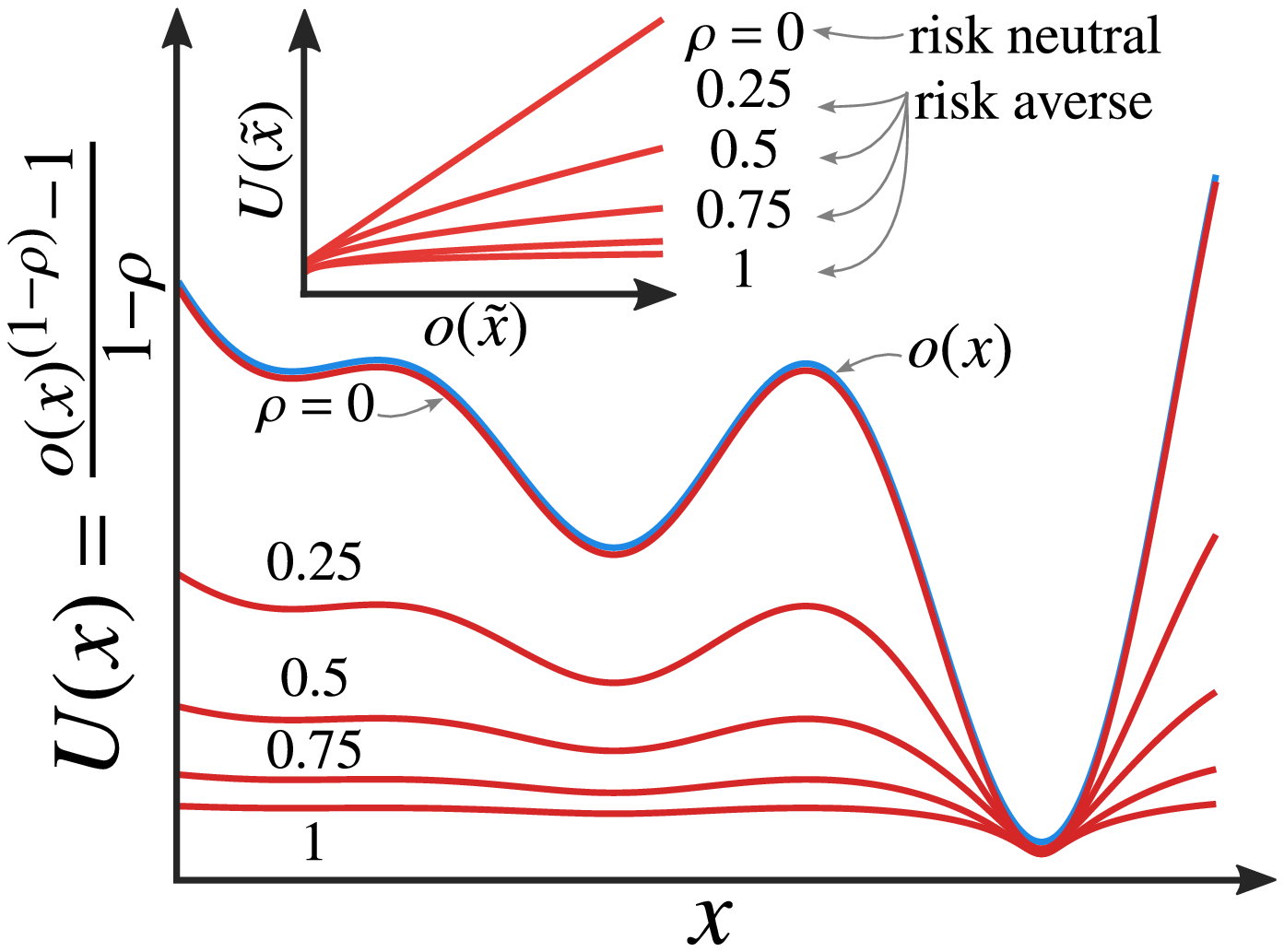}
\caption{Utility functions.}
\label{subfig:ufunc}
\end{subfigure}%
\begin{subfigure}[t]{0.25\textwidth}
\centering
\includegraphics[scale=0.33]{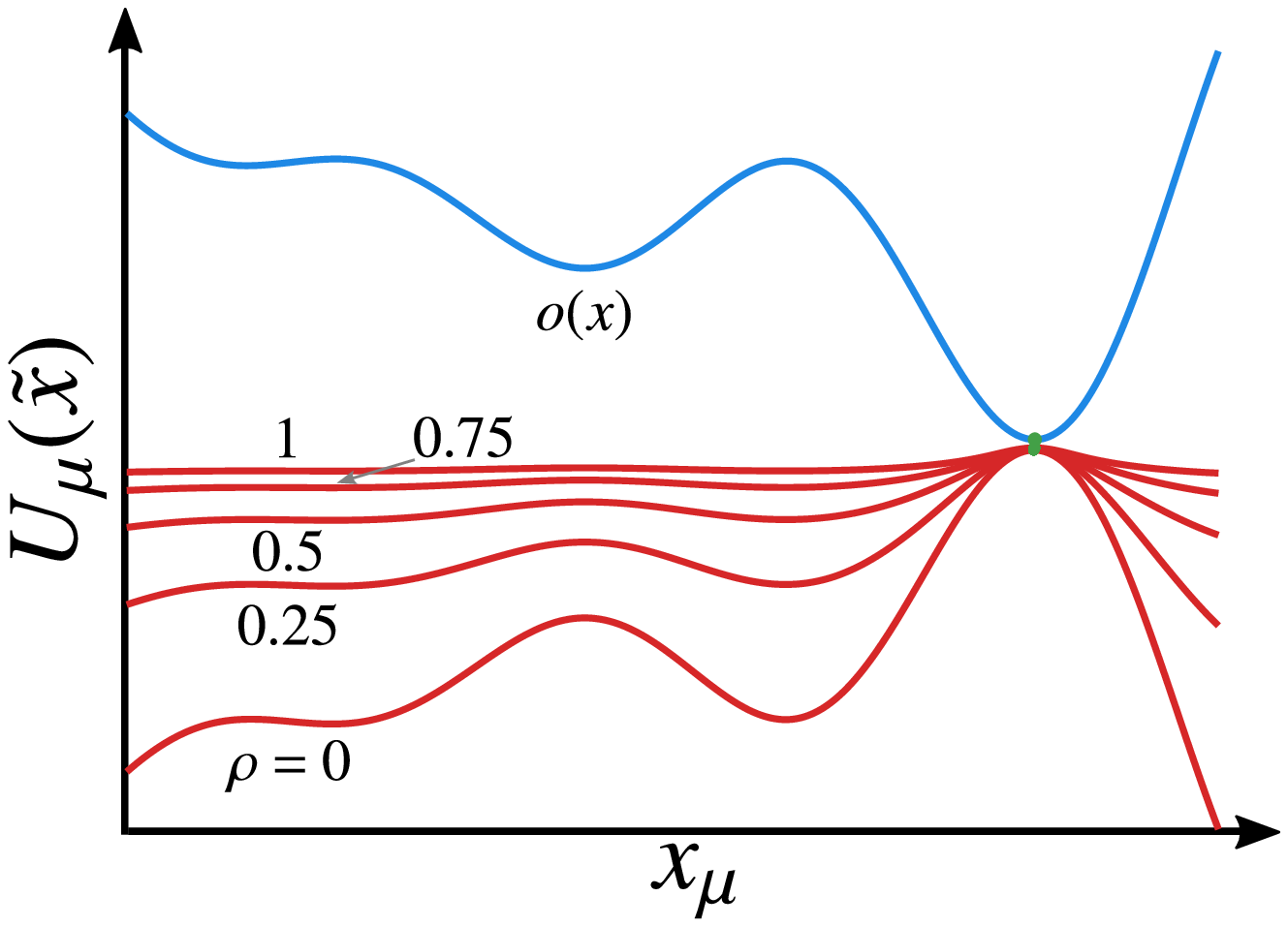}
\caption{Small uncertainty.}
\label{subfig:ul}
\end{subfigure}%
\begin{subfigure}[t]{0.25\textwidth}
\centering
\includegraphics[scale=0.33]{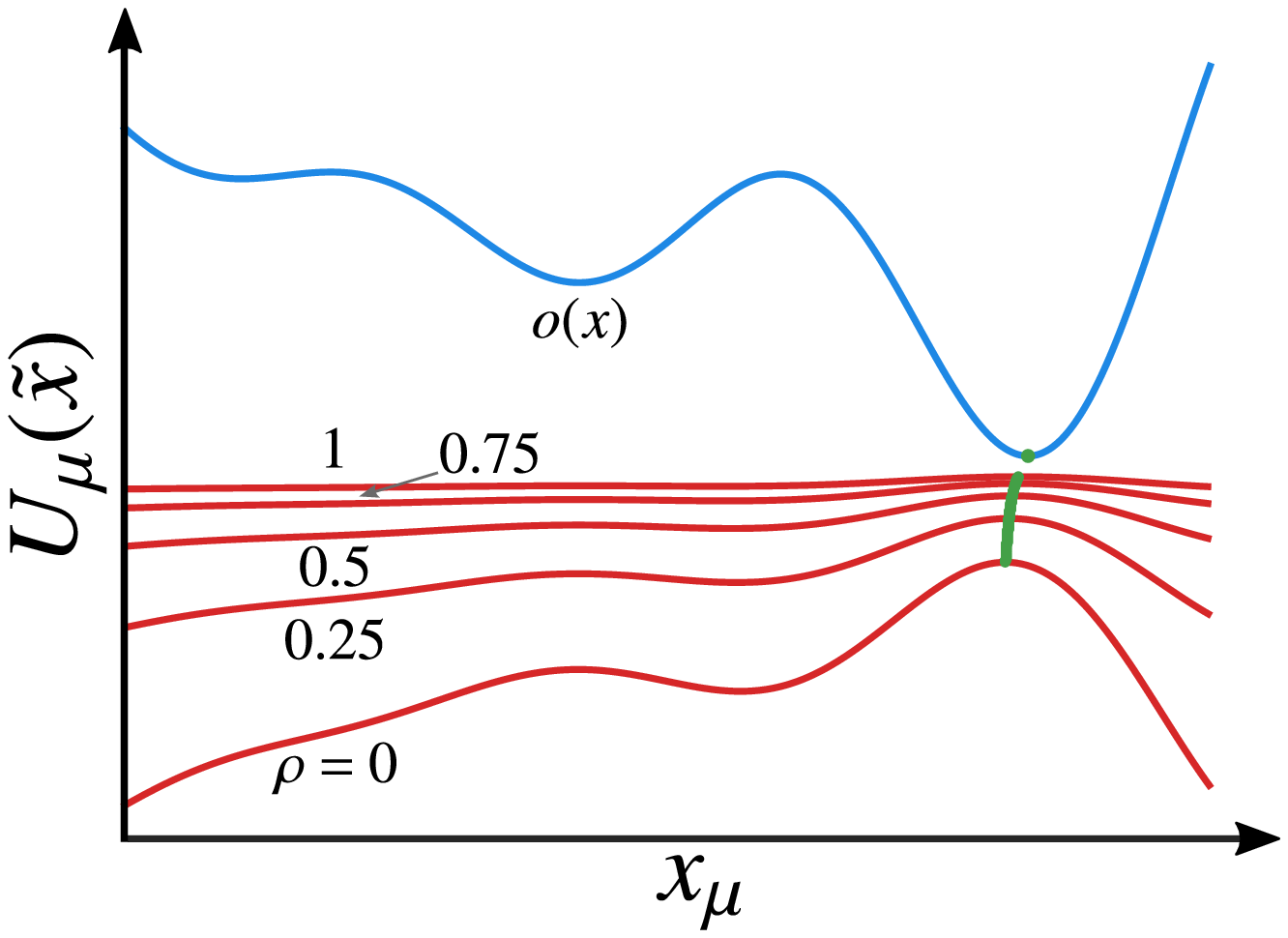}
\caption{Medium uncertainty.}
\label{subfig:um}
\end{subfigure}%
\begin{subfigure}[t]{0.25\textwidth}
\centering
\includegraphics[scale=0.33]{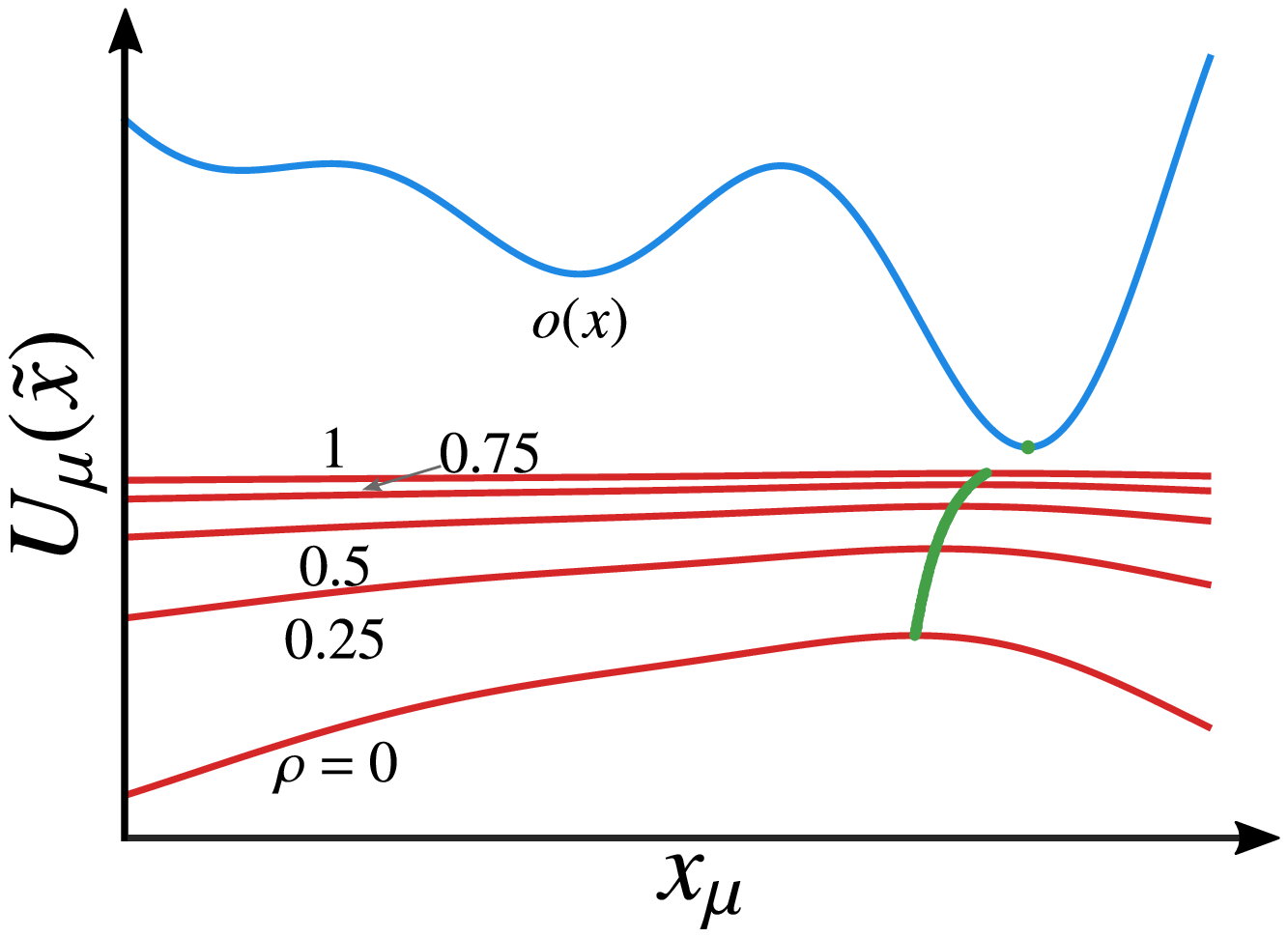}
\caption{Large uncertainty.}
\label{subfig:uh}
\end{subfigure}
\captionsetup[figure]{justification=centering}
\caption{Illustration of a constant relative risk averse utility function, $U_{\mu}(\tilde{x}) = -\mathbb{E}[\frac{o(\tilde{x})^{(1-\rho)-1}}{1-\rho}]$ with various relative risk aversion levels $\rho$ for three levels of uncertainty ($x_\sigma$) of a single uncertain variable $\tilde{x} = \mathcal{N}(x_{\mu},x^2_{\sigma})$.}
\label{figs:Utility}
\end{figure*}

For an arbitrary objective function, the probabilistic robust interpretation, along with the Pareto optimal front between the expectancy and dispersion terms for notional small, medium, and large uncertainties are presented in Fig.~\ref{fig:PRObjective}.
From Fig.~\ref{subfig:PRo_S}, it is clear that PR-UCCD is not always a risk-averse formulation because the optimal, multiobjective solution is invariant with respect to variance for the majority of weighting factors.
However, as uncertainties increase in size, the objective function exhibits more deviations compared to the deterministic case.
While, in Fig.~\ref{subfig:PRo_S}, this behavior is attributed to the magnitude of uncertainties, studies have shown that the usage of variance as a measure to quantify robustness has some limitations and requires restrictive assumptions \cite{malak2015decision,baron1977utility,bigelow1993consistency,basak2010dynamic}.
For instance, Malak et al. argue that using variance to quantify robustness can bias decision-makers toward demonstrably riskier alternatives, for example, when the underlying distributions have nonzero skew \cite{malak2015decision}.

% new paragraph
To address such limitations, one approach is to use concepts from normative decision theory, such as representation theorems \cite{sep-rationality-normative-utility}, that often result in a mathematical description of decision-maker's preferences through a utility function.
The shape of this utility function conveys information about decision-maker's risk attitude. 
For example, a locally concave utility function corresponds to a risk-averse attitude; a linear utility function corresponds to a risk-neutral attitude; and a convex utility function corresponds to a risk-taking attitude \cite{malak2015decision}.

% new paragraph
The usage of expected utility theory for the arbitrary objective function of Fig.~\ref{fig:PRObjective}, is presented in Fig.~\ref{figs:Utility}. 
In this illustration, we define a constant relative risk-averse utility function as $U(\tilde{x}) = \frac{o(\tilde{x})^{(1-\rho)}-1}{1-\rho}$.
The relative degree of risk aversion in this utility function is the constant $\rho$; therefore, the changes in $o(\tilde{x})$ do not affect the decision-maker's attitude towards risk. 
From Fig.~\ref{subfig:ufunc}, it is notable that when $\rho$ is close to zero, the utility function tends to linearity (i.e.~risk neutral), while for larger values of $\rho$ the utility function becomes concave (i.e.~risk averse). 
Here, the increasingly risk-averse behavior of the decision-maker (as $\rho$ goes from $0$ to $1$) is modeled through utility functions with increasingly less extreme changes over the function domain.
In other words, as we become more risk averse, the loss incurred from possibly losing the lottery (i.e.~not being able to realize the best objective function) decreases. 
Figures~\ref{subfig:ul}-\ref{subfig:uh} present these utility functions for notional small, medium, and large uncertainties.

\subsection{Insights From Robust Control Theory}
\label{subsec:RCT}
Robust control theory is involved with the analysis and synthesis of controllers that can mitigate the impact of uncertainties on performance specifications and stability. 
In classical control theory, these performance specifications are described through frequency or time domain measures.
Various tools such as gain and phase margins~\cite{paraskevopoulos2017modern}, disk margins~\cite{seiler2020introduction}, $H_{2}$, $H_{\infty}$, and $\mu$-synthesis ~\cite{dullerud2013course} have been developed to address uncertainty-related challenges.

\begin{figure}[t]
\centering
\includegraphics[scale=0.5]{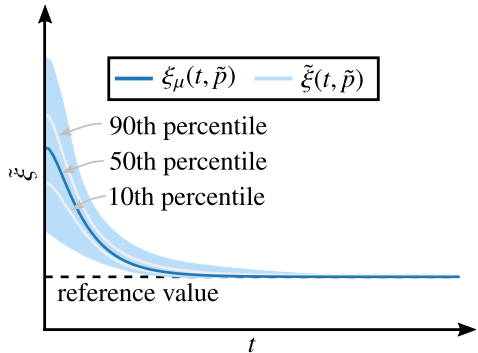}
\caption{Reference tracking of a stable control system with several uncertainties using an infinite-horizon linear quadratic regulator (LQR).}
\label{Figlqr}
\end{figure}

% new paragraph
The development of robust control theory has been largely dependent upon the benefits of feedback control. 
First, it should be emphasized that the generalized formulation introduced in Eq.~(\ref{Eqn:UCCD}) may entail control gains $\bm{p}_{c}$ that are used to establish a feedback control. 
In addition, while closed-loop control plays an essential role in mitigating the impact of some uncertainties in UCCD problems, these uncertainties still affect the dynamic system behavior and the overall system performance. 
As shown in 
Fig.~\ref{Figlqr}, a notional infinite-horizon linear quadratic regulator (LQR) (which is an optimal controller for its associated cost function) reduces uncertainty in the system response over time to the reference value, assuming stability under the uncertainties. 

% Figure 8

\subsection{Insights From Stochastic Control Theory}
\label{subsec:SCT}
In stochastic control theory, idealized processes such as stationary, normal, Markov, second-order, and Wiener are used to characterize the distribution of stochastic processes. 
Many of the disturbances affecting the control system can be modeled by processes generated from Wiener processes~\cite{aastrom2012introduction}.
While we previously assumed that the noise vector is included in the vector of problem data $\tilde{\bm{d}}$, to keep the notation consistent with stochastic control theory, here, we use $\tilde{\bm{w}}$ to describe an $n_{w}$-dimensional standard Brownian motion defined on a complete probability space.
The nonlinear stochastic system model can be described as:
\begin{align}
\label{Eqn:SCCD_Nonlin}
d\tilde{\bm{\xi}}(t) = \bm{f}(t, \tilde{\bm{u}},\tilde{\bm{\xi}}, \tilde{\bm{p}}, \tilde{\bm{d}})\,dt + \bm{b}(t, \tilde{\bm{u}}, \tilde{\bm{\xi}}, \tilde{\bm{p}}, \tilde{\bm{d}} )\,d\tilde{\bm{w}}(t) 
\end{align}
\noindent
where the $\bm{f}(\cdot)$ and $\bm{b}(\cdot)$ are maps that are commonly referred to as the drift and diffusion terms, respectively~\cite{yong2020stochastic}. 
Because standard Brownian motion is not differentiable, its associated integral form is commonly used instead and requires I\^to, Stratonovich, or backward integral approaches.

% new paragraph
A special case of Eq.~(\ref{Eqn:SCCD_Nonlin}) is when the dynamics are linear and the objective function is quadratic in $(\tilde{\bm{\xi}}(t), \tilde{\bm{u}}(t))$. 
This problem, referred to as a stochastic linear-quadratic problem (SLQ-UCCD), is significant because the optimal control law can be synthesized into a feedback form of the optimal state, and the corresponding proportional coefficients may be specified through the associated Ricatti equation.
This unique control law is a combination of the Kalman filter and LQR.
Additionally, we note that for linear systems with additive white noise, several tools become available.
For example, using linear filters such as
the Wiener filter in the frequency domain and Kalman filters in the state-space domain, one can separate the noise from the signal of interest by minimizing the mean-square error~\cite{cottrill2012hybrid}. 
Finally, there are other cases studied when the state equation is linear~\cite{yong2020stochastic, sun2020stochastic, chen2001stochastic}.

\subsection{Open-loop Control Structure Under Uncertainties}
There is an essential question on the role of optimal control trajectories in the open-loop formulation of UCCD problems.
In response to uncertainties, one may use an open-loop single-control (OLSC) or an open-loop multiple-control (OLMC) structure.
OLSC is structured to find a single control command, which is often used for reference tracking applications, while OLMC elicits a range of optimal control responses based on the realization of uncertainties.
Distinctions between the two structures are best manifested when solving boundary-value UCCD problems.
This is because, unlike OLMC, the single control command in OLSC cannot satisfy all of the prescribed initial and terminal boundary conditions in the presence of uncertainties.

% new paragraph
This issue has been dealt with in two different ways in the literature: (i) relaxing the prescribed terminal boundary conditions~\cite{azad2020single,  azad2021robust}, or (ii) minimizing the variance of the terminal state in a multi-objective optimization problem~\cite{fisher2011optimal, boutselis2016stochastic}.
These remedies enable a solution to the OLSC-UCCD problem, but they have limitations because they do not enforce the terminal boundary conditions.
This caveat is problematic because relaxing the boundary conditions is not practically viable for many real-world applications. 
Therefore, OLSC should be used selectively in the appropriate context.  

% new paragraph
On the other hand, OLMC is based on the idea that uncertainty realizations should elicit a distinct optimal control response from the UCCD problem (which has conceptual similarities to how closed-loop systems respond).
Because each distinct optimal control response is only associated with a specific uncertainty realization, all the initial and terminal boundary conditions may be satisfied in this control structure. 
Through this, OLMC provides additional insights into the uncertainty-informed limits of the system performance. 
Therefore, OLMC is suitable during early-stage design, where plant and control spaces are being explored, not only for optimal performance but also for reliability, robustness, or any other risk measures. OLSC and OLMC structures are compared in Ref.~\cite{Azad2022b}.

\subsection{Stochastic and Robust Model Predictive Control}
While all formulations introduced in this article consider a single-horizon UCCD problem, model predictive control (MPC) solves a sequence of such problems to find a cost-minimizing control action for a relatively short horizon in the future.
For online implementations, this controller has the advantage of the current state information to predict state trajectories that emanate from the current state.
The issue of uncertainties considered with robust and stochastic MPC~\cite{bemporad1999robust, mesbah2016stochastic, nash2021robust}.  

\subsection{Moving Forward}\label{subsec:MovFor}
Addressing uncertainty-related challenges in various engineering domains requires identifying ways to characterize that domain's uncertainties.
A critical aspect of this task is understanding the availability of information at different stages of the design process.
This, along with the risks involved in the specific application of interest, to some degree inform the designer's decision to employ any of the formulations presented in this article. 
Additionally, the choice of solution strategies is of great theoretical and practical importance. It requires consideration of several factors, including the suitability of the approach for specific designer preferences, domains, computational cost, accuracy, convergence, etc.
The impact of solution strategies on the integrated UCCD solution and their inclusion in various coordination strategies must be further investigated in order to provide answers to issues such as scalability.
An initial effort is Ref.~\cite{Azad2022b}, which offers some preliminary insights into comparisons between an SE-UCCD and WCR-UCCD using MCS and generalized polynomial chaos expansion.

%--------------Conclusions-----------------%
\xsection{Conclusion}\label{sec:conclusion}
With all the recent advances and applications of (deterministic) control co-design, significant work is still needed to handle uncertainty when developing effective combined plant and control solutions.
Investigating the current state-of-the-art for uncertain control co-design (UCCD), we have identified several significant assumptions.
Generally, the scope of uncertainties is limited to a single discipline (often either with a plant or control or even solution method emphasis).
Additionally, different interpretations and representations of uncertainty affect different problem elements, including the objective function, equality/inequality constraints, and optimization variables.

% new paragraph
To start to address these shortcomings, this article discussed a broad range of relevant uncertainties and the multitude of ways to characterize UCCD problem elements.
The discussion naturally led to six specialized UCCD problem formulations, including stochastic in expectation, stochastic chance-constrained, probabilistic robust, worst-case (minimax) robust, fuzzy expectation, and possibilistic chance-constrained.
These formulations are not disconnected; the link between minimax robust and stochastic chance-constrained UCCD was also discussed.

% new paragraph
Overall, this article aims at providing a concrete framework to discuss and represent uncertainties in UCCD, providing a foundation for additional advances, both in theory and applications of UCCD.
Understanding how to represent and interpret a domain's uncertainties is one of the first challenges.
A natural next step is to investigate methods and solution strategies corresponding to these formulations, seeking to balance various design goals and computational expense.

%--------------acknowledgment-----------------%
\begin{acknowledgment}
This research was partially supported by the National Science Foundation, Division of Civil, Mechanical, \& Manufacturing Innovation, Engineering Design and System Engineering Program, under grant number CMMI-2034040.
\end{acknowledgment}

%--------------References-----------------%
\renewcommand{\refname}{REFERENCES}
\bibliographystyle{asmems4}
{\small
\bibliography{References}
}

%--------------Nomenclature-----------------%
% Acronyms
\nomenclature[A, 01]{\(\textrm{AAO}\)}{all-at-once}
\nomenclature[A, 02]{a.s.}{almost surely}
\nomenclature[A, 03]{\(\textrm{CCD}\)}{control co-design}
\nomenclature[A, 04]{\(\textrm{CVaR}\)}{conditional value at risk }
\nomenclature[A, 05]{\(\textrm{FDE}\)}{fuzzy differential equations }
\nomenclature[A, 06]{FE}{fuzzy expected value}
\nomenclature[A, 07]{LQR}{linear quadratic regulator}
\nomenclature[A, 08]{MCS}{Monte Carlo simulation}
\nomenclature[A, 09]{MPC}{model predictive control}
\nomenclature[A, 10]{OLMC}{open-loop multiple control}
\nomenclature[A, 11]{OLSC}{open-loop single control}
\nomenclature[A, 12]{\(\textrm{ODE}\)}{ordinary differential equations }
\nomenclature[A, 13]{PCC}{possibilistic chance-constrained}
\nomenclature[A, 14]{PR}{probabilistic robust}
\nomenclature[A, 15]{RBDO}{reliability-based design optimization}
\nomenclature[A, 16]{RC}{robust counterpart}
\nomenclature[A, 17]{SCC}{stochastic chance-constrained}
\nomenclature[A, 18]{\(\textrm{SDE}\)}{stochastic differential equations }
\nomenclature[A, 19]{SE}{stochastic in expectation}
\nomenclature[A, 20]{\(\textrm{UCCD}\)}{uncertain control co-design}
\nomenclature[A, 21]{WCR}{worst-case robust}

% Key variable, functions, sets, etc.

% Capital and lower case Letters
\nomenclature[v, 01]{\( \mathcal{D}\)}{set of deterministic variables}
\nomenclature[v, 02]{\( \bm{d} \)}{vector of problem data}
\nomenclature[v, 03]{\( \tilde{\bm{d}} \)}{vector of uncertain problem data}
\nomenclature[v, 04]{\(\mathbb{E}\)}{expected value}
\nomenclature[v, 05]{\( \mathcal{E} \)}{ feasible set of Type~I equality constraints}
\nomenclature[v, 06]{\(\bm{f}(\cdot)\)}{state derivative function}
\nomenclature[v, 07]{\(\bm{g}(\cdot)\)}{inequality constraint vector}
\nomenclature[v, 08]{\(\bm{h}(\cdot)\)}{equality constraint vector}
\nomenclature[v, 09]{\( \mathbb{I}\)}{indicator function}
\nomenclature[v, 10]{\( \mathcal{I}_{D}\)}{constraint feasible set}
\nomenclature[v, 11]{\( \mathcal{I}_{RC} \)}{feasible space of the RC problem}
\nomenclature[v, 12]{\( k_{s} \)}{constraint shift index}
\nomenclature[v, 13]{\(\ell(\cdot)\)}{Lagrange term}
\nomenclature[v, 14]{\(M(\cdot)\)}{set membership function }
\nomenclature[v, 15]{\(m(\cdot)\)}{Mayer term}
\nomenclature[v, 17]{\( n \)}{number of elements in a set }
\nomenclature[v, 18]{\(o(\cdot)\)}{objective function}
\nomenclature[v, 19]{\( \mathbb{P}\)}{probability measure}
\nomenclature[v, 20]{\(\bm{p}\)}{vector of time-independent optimization variables}
\nomenclature[v, 21]{\( \tilde{\bm{p}}\)}{time-independent uncertain optimization variables}
\nomenclature[v, 22]{\( \mathbb{P}\mathbb{O}\mathbb{S}_{f,i} \)}{failure possibility for the $i$th constraint}
\nomenclature[v, 23]{\( \mathbb{P}_{f,i}\)}{target failure probability of $i$th constraint}
\nomenclature[v, 24]{\( \mathbb{P}_{f,sys}\)}{system target failure level }
\nomenclature[v, 25]{\(\bm{p}_{c}\)}{time-independent control optimization variables}
\nomenclature[v, 26]{\(\bm{p}_{p}\)}{time-independent plant optimization variables}
\nomenclature[v, 27]{\( \hat{\bm{q}} \)}{vector of nominal variables}
\nomenclature[v, 28]{\( \mathcal{R} \)}{uncertainty set used in WCR}
\nomenclature[v, 29]{\(t\)}{time}
\nomenclature[v, 30]{\( \mathcal{U}\)}{set of uncertain variables}
\nomenclature[v, 31]{\(\bm{u}\)}{control vector}
\nomenclature[v, 32]{\(\tilde{\bm{u}}\)}{uncertain control vector}
\nomenclature[v, 33]{\( v \)}{objective function variable in epigraph form}
\nomenclature[v, 34]{\(\tilde{\bm{w}}\)}{uncertain noise vector}
\nomenclature[v, 35]{\( \mathcal{X}_{\text{crisp}} \)}{ crisp description of uncertainty set}
\nomenclature[v, 36]{\( \mathcal{X}_{\text{fuzzy}} \)}{ fuzzy description of uncertainty set}
\nomenclature[v, 37]{\( \mathcal{X}_{\text{stc}} \) }{stochastic description of uncertainty set}
\nomenclature[v, 38]{\(x\)}{an arbitrary deterministic variable}

% Greek letters 
\nomenclature[v, 40]{\( \alpha_{w} \)}{weighting factor}
\nomenclature[v, 41]{\(\beta_t\)}{target reliability level }
\nomenclature[v, 43]{\(\bm{\xi}\)}{state vector} 
\nomenclature[v, 44]{\(\tilde{\bm{\xi}}\)}{uncertain state vector} 
\nomenclature[v, 45]{\( \bm{\sigma}_{a} \)}{allowable standard deviation vector associated with $\bm{g}$}
\nomenclature[v, 46]{\(\Phi(\cdot)\)}{constraint maximization problem in Eq.~(\ref{Eqn:RCCDc})}
\nomenclature[v, 47]{\( \psi(\cdot) \)}{feasibility constraint in Eq.~(\ref{Eqn:RCCDc})}

% Key subscripts
\nomenclature[B, 01]{\( 0 \)}{initial}
\nomenclature[B, 02]{\( N \)}{nominal values}
\nomenclature[B, 03]{\( f \)}{final}
\nomenclature[B, 04]{\( i \)}{counter}
\nomenclature[B, 05]{\( t \)}{time-dependent}
\nomenclature[B, 06]{\( \mu \)}{mean-value }
\nomenclature[B, 07]{\( \sigma \)}{standard deviation}

% Needs to be updated
\mbox{}

\printnomenclature

\end{document}